\pdfoutput=1
\documentclass[journal=enfuem,manuscript=article,hyperref]{achemso}

\usepackage{hyperref}
\usepackage{achemso}
\usepackage[version=3]{mhchem} % Formula subscripts using \ce{}, e.g., \ce{H2SO4}
\usepackage{amsmath,amssymb}
\usepackage{bm}

\usepackage{booktabs,multicol,multirow}

\usepackage{graphicx}
\usepackage{caption}
\usepackage{subcaption}

\usepackage[english]{babel}
\usepackage{textcomp}

\usepackage[version-1-compatibility]{siunitx}
\DeclareSIUnit\atm{atm}

\usepackage{algorithm,algorithmic}
\usepackage{url,doi}

\usepackage{setspace}

\DeclareMathOperator{\sgn}{sgn}

\makeatletter
\newcounter{reaction}
\renewcommand\thereaction{C\,\arabic{reaction}}
\newcommand\reactiontag%
  {\refstepcounter{reaction}\tag{\thereaction}}
\newcommand\reaction@[2][]%
  {\begin{equation}\ce{#2}%
  \ifx\@empty#1\@empty\else\label{#1}\fi%
  \reactiontag\end{equation}}
\newcommand\reaction@nonumber[1]%
  {\begin{equation*}\ce{#1}\end{equation*}}
\newcommand\reaction%
  {\@ifstar{\reaction@nonumber}{\reaction@}}
\makeatother

\author{Kyle E.\ Niemeyer}
\email{Kyle.Niemeyer@oregonstate.edu}
\affiliation[Oregon State University]
{School of Mechanical, Industrial, and Manufacturing Engineering, Oregon State University, Corvallis, OR, USA}

\author{Chih-Jen Sung}
\affiliation[University of Connecticut]
{Department of Mechanical Engineering, University of Connecticut, Storrs, CT, USA}

\title[]
{Reduced chemistry for a gasoline surrogate valid at engine-relevant conditions}

\begin{document}
\singlespacing

%%%%%%%%%%%%%%%%%%%%%%%%%%%%%%%%%%%%%%%%%%%%%%%%%%%%%%%%%%%%%%%%%%%%%
%% The manuscript does not need to include \maketitle, which is
%% executed automatically.  The document should begin with an
%% abstract, if appropriate.  If one is given and should not be, the
%% contents will be gobbled.
%%%%%%%%%%%%%%%%%%%%%%%%%%%%%%%%%%%%%%%%%%%%%%%%%%%%%%%%%%%%%%%%%%%%%
\begin{abstract}
A detailed mechanism for the four-component RD387 gasoline surrogate developed by Lawrence Livermore National Laboratory has shown good agreement with experiments in engine-relevant conditions.
However, with 1388 species and 5933 reversible reactions, this detailed mechanism is far too large to use in practical engine simulations.
Therefore, reduction of the detailed mechanism was performed using a multi-stage approach consisting of the DRGEPSA method, unimportant reaction elimination, isomer lumping, and analytic QSS reduction based on CSP analysis.
A new greedy sensitivity analysis algorithm was developed and demonstrated to be capable of removing more species for the same error limit compared to the conventional sensitivity analysis used in DRG-based skeletal reduction methods.
Using this new greedy algorithm, several skeletal and reduced mechanisms were developed at varying levels of complexity and for different target condition ranges.
The final skeletal and reduced mechanisms consisted of 213 and 148 species, respectively, for a lean-to-stoichiometric, low-temperature HCCI-like range of conditions.
For a lean-to-rich, high-temperature, SI\slash CI-like range of conditions, skeletal and reduced mechanisms were developed with 97 and 79 species, respectively.
The skeletal and reduced mechanisms in this study were produced using an error limit of \SI{10}{\percent} and validated using homogeneous autoignition simulations over engine-relevant conditions---all showed good agreement in predicting ignition delay.
Furthermore, extended validation was performed, including comparison of autoignition temperature profiles, PSR temperature response curves and extinction turning points, and laminar flame speed calculations.
All the extended validation showed results within\slash near the \SI{10}{\percent} error limit, demonstrating the adequacy of the resulting reduced chemistry.
\end{abstract}

\section{Introduction}
\label{S:intro}
%%%%%%%%%%%%%%%%%%%%%%%%%%%%%%%%%%%%%%%%%%%%%%%%%%%%%%%%%%%%%%%%%%%%%

Modeling the kinetics of gasoline---as well as other liquid transportation fuels---is complex due to the near-continuous spectrum of constituent hydrocarbons.
One widely used solution in the combustion community is to use surrogate fuels that consist of a small number of hydrocarbons representing the major hydrocarbon classes present in real gasoline.
Historically, binary blends of \emph{n}-heptane and isooctane were used to model gasoline at various octane numbers; these are the primary reference fuels (PRFs)\cite{Edgar:1927fq,Sturgis:1954uq,Pahnke:1954er}.
However, in general these simple mixtures cannot match some key properties of gasoline.
For example, the H\slash C ratio of gasoline is usually less than two~\cite{Chaos:2007}, but PRFs are limited to the range of \numrange{2.3}{2.25}.
In addition, PRFs cannot capture the so-called gasoline sensitivity, the difference between motor octane number (MON) and research octane number (RON); RON and MON are equal for any PRF mixture.

In order to better match the physical and chemical properties of gasoline, a number of research groups developed surrogate formulations containing additional components used to represent other major hydrocarbon classes (e.g., olefins, aromatics).
Gauthier et al.~\cite{Gauthier:2004} and Chaos et al.~\cite{Chaos:2007} proposed three-component surrogates, adding toluene to \emph{n}-heptane and isooctane to form toluene reference fuels (TRFs).
Recently, Mehl et al.~\cite{Mehl:2011cn,Mehl:2011jn} proposed a four-component surrogate for RD387 gasoline  consisting of \emph{n}-heptane, isooctane, toluene, and 2-pentene to represent linear alkanes, branched alkanes, aromatics, and olefins, respectively.
They found that this surrogate emulates engine data, laminar flame speeds, and shock tube ignition delay times of the target gasoline with good agreement.
Kukkadapu et al.~\cite{Kukkadapu:2012dx,Kukkadapu:2013ko} performed further experimental and computational validation of the surrogate mixture and representative kinetic mechanism of Mehl et al.~\cite{Mehl:2011cn}.
They found that for stoichiometric mixtures the surrogate matched the autoignition response of the target gasoline in a rapid compression machine, and the mechanism predicted overall ignition delays of real gasoline with good agreement.
Sarathy et al.~\cite{Sarathy:2014aa} later used the methodology of Mehl et al.~\cite{Mehl:2011jn} to create multicomponent surrogates for alkane-rich FACE (Fuels for Advanced Combustion Engines) gasoline fuels.

While the performance of the proposed RD387 gasoline surrogate mechanism is promising, the large size of the reaction mechanism---1388 species and 5933 reversible reactions---poses a significant challenge to practical engine simulations.
The computational cost of chemistry scales by the third power of the number of species in the worst case~\cite{Lu:2009gh}.
In three-dimensional, high-fidelity simulations of engines or combustion chambers where mesh sizes could range \numrange{e4}{e7} cells, chemistry calculations must be performed at least once for each grid point or cell.
Therefore, significant reduction in mechanism size while retaining its predictive capabilities is vital in order to use the mechanism in practical simulations.
With this in mind, the objectives of the current study are to 
\begin{enumerate}
\item develop and demonstrate a multi-stage mechanism reduction methodology capable of achieving the above task, and
\item produce compact skeletal and reduced mechanisms for the aforementioned RD387 gasoline surrogate capable of predicting key combustion phenomena.
\end{enumerate}
While the reduced chemical models demonstrated here are specifically applicable only to the RD387 gasoline surrogate, the mechanism reduction approach will be useful in general for reduction of large detailed mechanisms for multicomponent surrogate fuels.

A number of mechanism reduction methods have been developed in recent years to counter the trend of increasing mechanism sizes, as reviewed by Lu and Law~\cite{Lu:2009gh}.
Most approaches focus on identifying and removing unimportant species, or performing ``skeletal'' reduction.
Many methods have been developed, but one class that received significant development is based on the directed relation graph (DRG)~\cite{Lu:2005ce,Lu:2006bb,Lu:2006gi}.
Similar to the earlier graphical representation of reaction pathways of Bendtsen et al.~\cite{Bendtsen:2001vh}, DRG quantifies the importance of species using normalized contributions to the overall production rates of certain (preselected) important target species.
Since the introduction of DRG, a number of variants have been developed, including DRG-aided sensitivity analysis (DRGASA)~\cite{Sankaran:2007fs,Zheng:2007gd,Lu:2008bi}, DRG with error propagation (DRGEP)~\cite{Pepiot-Desjardins:2008,Niemeyer:2011fe}, DRGEP with sensitivity analysis (DRGEPSA)~\cite{Niemeyer:2010bt,Niemeyer:2014}, and path-flux analysis~\cite{Sun:2010jf}.

Another reduction paradigm focuses on time-scale analysis, identifying and removing short times scales---induced by rapidly depleting species and/or fast reversible reactions---that cause chemical stiffness.
Many methods rely on the classical quasi-steady state (QSS)~\cite{Bodenstein:1913tc,Chapman:1913dx} and partial equilibrium approximations~\cite{Benson:1952ju,Ramshaw:1980kn}, which replace differential equations with algebraic relations for some species.
Originally, such species and reactions were identified on the basis of experience and intuition, but systematic methods that use analysis of the Jacobian matrix to identify QSS species and partial equilibrium reactions, namely the computational singular perturbation (CSP)~\cite{Lam:1988wc,Lam:1993ub,Lam:1994ws} and intrinsic low-dimensional manifold~\cite{Maas:1992ws} methods, have since been developed.

Finding a single skeletal reduction stage not sufficient to reduce the size of large detailed mechanisms, Lu and Law~\cite{Lu:2008bi} presented a multi-stage reduction strategy and applied it to to a detailed mechanism for \emph{n}-heptane.
Their approach consisted of DRGASA, unimportant reaction elimination, isomer lumping, and time scale reduction through the QSS approximation.
In addition, they grouped species with similar transport properties in the final reduced mechanism to reduce the cost of the mixture-averaged transport formulation.

In this work, we apply a multi-stage reduction strategy similar to that developed by Lu and Law~\cite{Lu:2008bi} based on DRGEPSA to the large detailed mechanism for the RD387 gasoline surrogate of Mehl et al.~\cite{Mehl:2011cn,Mehl:2011jn}.
We selected DRGEPSA for the base skeletal reduction method following the demonstration of Niemeyer et al.~\cite{Niemeyer:2010bt} that DRGEPSA can produce more compact skeletal mechanisms for the same level of accuracy than DRG, DRGEP, and DRGASA.
In the first stage, the DRGEPSA method is applied to remove a large number of unimportant species (and corresponding reactions).
Second, a stage of further unimportant reaction elimination is performed to remove additional reactions; this step does not affect the number of species, but the complexity of the mechanism is reduced.
Third, a final skeletal mechanism is produced after identifying and lumping isomers.
Finally, QSS species are identified using CSP analysis and an analytic solution for the QSS species concentrations is generated.
We then validate the resulting skeletal and reduced mechanisms over engine-relevant conditions, comparing the performance against that of the detailed mechanism in predicting global phenomena such as homogeneous, adiabatic ignition delays, perfectly stirred reactor extinction turning points, and laminar flame speeds; in addition, comparisons between local temperature evolution profiles in autoignition provide more rigorous validation.

In the following sections, we will first describe the above multi-stage reduction methodology.
A new greedy sensitivity analysis algorithm will be introduced and developed that can achieve greater reduction than the conventional algorithm for the same error limit.
Then, skeletal and reduced mechanisms at varying levels of complexity will be generated using our reduction approach.
Finally, validation of the skeletal and reduced mechanisms will be performed, followed by a discussion of these results.

%====================================================================
\section{Methodologies}

In the current work, we applied a multi-stage reduction procedure that consisted of a number of previously developed techniques---with a number of updates and improvements---to the detailed mechanism for the gasoline surrogate of Mehl et al.~\cite{Mehl:2011cn,Mehl:2011jn}.
The mechanism, developed to represent a four-component surrogate (isooctane, \emph{n}-heptane, toluene, and 2-pentene) of RD387 gasoline, consists of 1388 species and 5933 reversible reactions.
As discussed by Niemeyer and Sung previously~\cite{Niemeyer:2014}, the current version of this mechanism contains a dead-end pathway terminating with the \ce{nC4H3} radical; as such, this species and the two reactions in which it participates were removed prior to mechanism reduction and later validation.
The RD387 gasoline surrogate formulation consists of \SI{48.8}{\percent} isooctane, \SI{15.3}{\percent} \emph{n}-heptane, \SI{30.6}{\percent} toluene, and \SI{5.3}{\percent} 2-pentene (by molar percentage)~\cite{Mehl:2011cn}, denoted as the LLNL surrogate hereafter.

Note that because the particular LLNL surrogate formulation was chosen for the mechanism reduction, the resulting skeletal and reduced mechanisms are only guaranteed to perform well for this mixture composition.
As recently discussed by Niemeyer and Sung~\cite{Niemeyer:2014}, using skeletal mechanisms outside the target mixture composition range can result in large errors.
Recognizing this, in the current study we checked the performance of the resulting skeletal mechanisms with varying mixture composition.
While better performance can be obtained outside the target range by adding additional mixtures or the neat fuels to the set of input conditions, this comes at the cost of a (potentially significantly) larger skeletal mechanism.
Niemeyer and Sung~\cite{Niemeyer:2014} also demonstrated that when performing mechanism reduction for multicomponent fuels, a greater extent of reduction can be achieved by using the mixture as input for a surrogate reduction rather than combining independent skeletal mechanisms produced for each neat component.
Another important finding of this study was that such a combination may overlook important interactions between the components~\cite{Niemeyer:2014}.
The surrogate reduction strategy therefore forms the basis of the current work.

Two different ranges of conditions were used: one targeted at low-temperature homogenous charge compression ignition (HCCI) engine conditions, and the second targeted at more traditional spark- or compression-ignition (SI or CI) engine conditions at higher temperatures.
Thermochemical data for the HCCI-like conditions were generated using constant-volume autoignition simulations performed using the initial conditions listed in \ref{T:conditions}, based on a similar set used by Mehl et al.~\cite{Mehl:2011jn}.
These include a range of conditions covering initial pressures of \SIrange{10}{60}{\atm}, initial temperatures of \SIrange{750}{1200}{\kelvin}, and equivalence ratios of \numrange{0.2}{1.0}.

For the SI\slash CI engine conditions, we used autoignition initial conditions covering \SIrange{1000}{1400}{\kelvin}, \SIrange{1}{40}{\atm}, and equivalence ratios of \numrange{0.5}{1.5}.
In addition, since the target phenomena include high-temperature flame propagation, for this range of conditions we performed the reduction including perfectly stirred reactor (PSR) data---in addition to that from autoignition---covering the same range of pressures and equivalence ratios with an inlet temperature of \SI{300}{\kelvin}.

Autoignition data were sampled densely during the ignition evolution, as described previously~\cite{Niemeyer:2010bt,Niemeyer:2010,Niemeyer:2014}.
PSR data were sampled at three points along the upper stable branch of the temperature response curve: (1) at the extinction turning point, (2) the point closest to \SI{0.1}{\second}, and (3) the logarithmic midpoint between points one and two.\footnote{The logarithmic midpoint, or geometric mean, is equal to the square root of the product of the surrounding points, i.e., $\tau_2 = \sqrt{\tau_1 \tau_3}$.}
For both the HCCI and SI\slash CI reductions, we applied an error limit of \SI{10}{\percent}, corresponding to error in ignition delay for the autoignition simulations.
We note that the original detailed mechanism for the LLNL RD387 gasoline surrogate has been validated against a wide range of experimental data~\cite{Mehl:2011cn,Mehl:2011jn,Kukkadapu:2012dx,Kukkadapu:2013ko}, and therefore ensuring close agreement between the skeletal and reduced mechanisms and the detailed mechanism also ensures agreement with experimental data used for mechanism validation.
For the PSR simulations, the error corresponded to the maximum error in predictions of residence time at the extinction turning point (point one) and the response temperature at points two and three.
For both sets of conditions, we selected isooctane, \emph{n}-heptane, toluene, 2-pentene, oxygen, and nitrogen (to prevent removal) as the DRGEP target species.

Both the autoignition and PSR simulations were parallelized using OpenMP~\cite{OpenMP:2008}, such that multiple simulations may be performed simultaneously.
This greatly reduced the runtime of the original sampling of the detailed mechanism in addition to that of the overall reduction process.

\subsection{Reduction procedure}

The overall reduction procedure consists of stages in two categories: the skeletal reduction stages, including the DRGEPSA method, unimportant reaction elimination, and isomer lumping; and the time-scale reduction stage, which includes CSP analysis, applying the QSS approximation, and generating the analytic QSS solution.
The details of these reduction stages are described in the following sections.

\subsubsection{Skeletal reduction stages}

First, we applied the DRGEPSA method, as described by Niemeyer et al.~\cite{Niemeyer:2010bt,Niemeyer:2010,Niemeyer:2014}, to remove a large number of species and reactions from the detailed mechanism.
We describe this method briefly here; further detail can be found in our prior work~\cite{Niemeyer:2010bt,Niemeyer:2010,Niemeyer:2014}.
The DRGEPSA method consists of using DRGEP~\cite{Pepiot-Desjardins:2008} first to quantify the importance of species to predetermined target species through a graph-based representation of species interdependence in the reaction system.
After forming the graph by quantifying the interdependence of all species pairs, a graph search is performed using Dijkstra's algorithm~\cite{Niemeyer:2011fe} initiating at the user-determined target species.
Species are then declared unimportant and removed when their importance value (overall importance coefficient, $R_{AB}$) falls below a cutoff threshold $\epsilon_{\text{EP}}$, which is determined iteratively based on the user-specified error limit.
Following the application of DRGEP, sensitivity analysis (SA) is performed on certain remaining ``limbo'' species based on their $R_{AB}$ values.
Species with overall importance coefficients above an upper threshold value ($\epsilon^*$) are automatically retained, while the limbo species, for which $\epsilon_{\text{EP}} \leq R_{AB} < \epsilon^*$, are considered individually for removal during the SA stage.

Previously used sensitivity analysis approaches removed limbo species one-by-one, arranged them in ascending order based on the error induced to the mechanism by their removal, and then removed the limbo species in this order until the global error reached a limit~\cite{Sankaran:2007fs,Zheng:2007gd,Lu:2008bi,Niemeyer:2010bt,Niemeyer:2010,Niemeyer:2014}.
We refer to this approach in the following as the ``initially informed'' sensitivity analysis algorithm.
In this work, we introduce an improved sensitivity analysis algorithm based on a novel ``greedy'' approach.
Greedy algorithms make locally optimal decisions with the goal of reaching global optima, using current information to make the choice that appears best~\cite{Cormen:2009uw}.
The new greedy sensitivity analysis algorithm first evaluates the error induced by the removal of each species, given by
\begin{equation}
	\delta_S = \left | \delta_{S, \text{ind}} - \delta_{ \text{skel} } \right | \;,
\label{E:SA-error}
\end{equation}
where $\delta_{\text{skel}}$ is the error of the current skeletal mechanism (prior to temporary removal of limbo species \emph{S}) and $\delta_{S, \text{ind}}$ the error induced by the removal of limbo species $S$, by removing each one-by-one.
Then, using the criterion given by Eq.~\eqref{E:SA-error}, the algorithm identifies the limbo species with the smallest induced error and removes it; this procedure is repeated until the maximum error reaches the user-specified limit.
By using $\delta_S$ rather than $\delta_{S, \text{ind}}$, the species whose removal affects the mechanism the least is selected for removal.
Admittedly, this sensitivity analysis algorithm is more computationally expensive due to the large number of error evaluations (and associated autoignition\slash PSR simulations) and would likely be prohibitive for direct application to a detailed mechanism.
However, the DRGEPSA approach makes this expense more tractable by using the DRGEP stage to initially eliminate a large number of species from the starting detailed mechanism.

In contrast, the previous initially informed sensitivity analysis algorithm~\cite{Sankaran:2007fs,Zheng:2007gd,Lu:2008bi,Niemeyer:2010bt} used only the species' original induced error information.
In that case, following the initial species removal the ordering of species is based on outdated information, leading to potentially suboptimal decisions for removal.
Therefore, the algorithm could exit early, resulting in a larger-than-necessary skeletal mechanism.
We will compare the results of the original initially informed and new greedy sensitivity analysis algorithms in due course.

Following the application of DRGEPSA, an additional step of further unimportant reaction elimination was performed based on the methodology of Lu and Law~\cite{Lu:2008bi}.
Again, additional details on the method can be found elsewhere~\cite{Lu:2008bi,Niemeyer:2014}.
This method uses the CSP importance index to quantify the importance of reactions to all species in the DRGEPSA-generated mechanism.
As with the DRGEP method, the cutoff threshold for reaction importance $\epsilon_{\text{reac}}$ was determined iteratively based on error limit.

Next, we applied an isomer lumping stage to further reduce the number of species~\cite{Lu:2008bi}.
Due to the similarity of many isomers' thermodynamic and transport properties, such species can be lumped together in the chemical kinetics and transport equations.
While more complex lumping strategies exist~\cite{Ahmed:2007fa,Pepiot:2008kq}, our procedure---following that of Lu and Law~\cite{Lu:2008bi}---is based on the observation that many isomers mass fractions are correlated to the mass fraction of their group (the sum of the isomers that comprise the group):
\begin{equation}
Y_{k,j} = \alpha_{k,j} Y_j \;,
\end{equation}
where $Y_{k,j}$ represents the mass fraction of isomer $k$ in group $j$, $Y_j$ is the total mass fraction of isomer group $j$, and $\alpha_{k,j}$ is a constant coefficient.
For each group of isomers we determined the constant $\alpha_{k,j}$ by sampling the isomer mass fractions using autoignition\slash PSR simulations and performing a linear regression.
To ensure the validity of data points, we only used reaction states where the isomer group mass fraction was non-negligible (i.e., where $Y_j \geq$\num{e-10})~\cite{Lu:2008bi}.
In addition, unlike previous efforts~\cite{Lu:2008bi} we automated the isomer selection by only choosing isomers for lumping when the coefficient of determination (i.e., $r^2$ value) was above a certain threshold $\epsilon_{\text{isom}}$, determined iteratively based on error.
Once the isomer coefficients were determined, the skeletal mechanism was lumped by modifying the rate coefficients of reactions where each isomer is a reactant with the coefficient $\alpha_{k,j}$, and replacing the isomer with a notional species for the group to which it belongs.

\subsubsection{Time-scale reduction stage}

The final reduction stage involved applying the QSS approximation to species identified using the CSP analysis of Lam and coworker~\cite{Lam:1988wc,Goussis:1992ez,Lam:1993ub,Lam:1994ws}, and follows a similar approach to that of Lu and Law~\cite{Lu:2008dp,Lu:2008bi}.
The reaction system can be represented with the ODEs
\begin{align}
\frac{d \mathbf{y}}{dt} &= \mathbf{g} ( \mathbf{y} ) \label{E:reac-ode} \;, \text{ and} \\
\frac{d \mathbf{g}}{dt} &= \mathbf{J} \mathbf{g} \;, \quad \mathbf{J} \equiv \frac{d \mathbf{g}}{d \mathbf{y}} \;,\label{E:jacobian}
\end{align}
where $\mathbf{y}$ is the species concentration vector, $\mathbf{g}$ is the species production rate vector, and $\mathbf{J}$ is the Jacobian matrix.
CSP analysis decomposes the source terms $\mathbf{g}$ into ``modes'' using basis vectors:
\begin{align}
\mathbf{f} &= \mathbf{B} \mathbf{g} \;, \\
\frac{d \mathbf{f}}{dt} &= \bm{\Lambda} \mathbf{f} \;, \\
\bm{\Lambda} &= \left( \frac{d \mathbf{B}}{dt} + \mathbf{B} \mathbf{J} \right) \mathbf{A} \;, \label{E:lambda-orig}\\
\mathbf{A} &= \mathbf{B}^{-1} \;,
\end{align}
where the matrices $\mathbf{A}$ and $\mathbf{B}$ hold the column and row basis vectors, respectively, and $\mathbf{f}$ is the vector of modes.
Practically, this procedure is implemented by calculating the Jacobian $\mathbf{J}$ using a sixth-order central finite difference, then using the LAPACK subroutine DGEEV~\cite{Anderson:1999} to calculate the eigenvalues and eigenvectors.
The Jacobian is assumed to be time independent such that the basis rotation term $\frac{d \mathbf{B}}{dt} = 0$, and as a result Eq.~\eqref{E:lambda-orig} becomes
\begin{equation}
\label{E:lambda-new}
\bm{\Lambda} = \mathbf{B} \mathbf{J} \mathbf{A} \;,
\end{equation}
or
\begin{equation}
\label{E:eigendecomp}
\mathbf{J} = \mathbf{A} \bm{\Lambda} \mathbf{B} \;,
\end{equation}
where the diagonal elements of $\bm{\Lambda}$ are the eigenvalues of $\mathbf{J}$, the columns of $\mathbf{A}$ contain the right eigenvectors, and the rows of $\mathbf{B}$ contain the left eigenvectors.

Next, the fast and slow subspaces are separated, such that
\begin{equation}
\frac{d}{dt} \binom{\mathbf{f}^{\text{fast}}}{\mathbf{f}^{\text{slow}}} = 
\begin{pmatrix}
\bm{\Lambda}^{\text{fast}} & \\
& \bm{\Lambda}^{\text{slow}}
\end{pmatrix}
\binom{\mathbf{f}^{\text{fast}}}{\mathbf{f}^{\text{slow}}} \;.
\end{equation}
The fast modes are those that rapidly exhaust and decay, while the slow modes remain important and control the overall behavior of the system.
The eigenvalues associated with fast modes---the diagonal elements of $\bm{\Lambda}^{\text{fast}}$---are negative with a much larger magnitude than the eigenvalues associated with the slow modes, contained in $\bm{\Lambda}^{\text{slow}}$.
The separation of the fast and slow subspaces is identified by a timescale analysis:
\begin{equation}
\frac{-1}{\lambda_{\min} \left( \bm{\Lambda}^{\text{fast}} \right)} \equiv \tau_{\text{fast}} < \frac{ \tau_\text{c} }{\alpha_{\text{CSP}}} \;,
\end{equation}
where the time scale of the fast subspace ($\tau_{\text{fast}}$) is the negative inverse of the smallest magnitude eigenvalue ($\lambda_{\min}$) in $\bm{\Lambda}^{\text{fast}}$, $\tau_{\text{c}}$ is a characteristic time scale of the reacting system (e.g., autoignition delay, extinction turning point residence time), and $\alpha_{\text{CSP}}$ is a safety factor (e.g., 100).

Once the fast and slow subspaces are separated, the species rates of production can be projected onto the two subspaces:
\begin{equation}
\mathbf{g} = \left( \mathbf{Q}^{\text{fast}} + \mathbf{Q}^{\text{slow}} \right) \mathbf{g} \;,
\end{equation}
where the fast and slow projection matrices are
\begin{align}
\mathbf{Q}^{\text{fast}} &= \mathbf{A}^{\text{fast}} \mathbf{B}^{\text{fast}} \;, \text{ and} \\
\mathbf{Q}^{\text{slow}} &= \mathbf{A}^{\text{slow}} \mathbf{B}^{\text{slow}} \;,
\end{align}
respectively.
The $i$th species is considered QSS if it satisfies the following condition over the entire parameter range of interest:
\begin{equation}
\left| \mathbf{Q}^{\text{slow}}_{\text{i,i}} \right| < \epsilon_{\text{CSP}} \;,
\end{equation}
where $ \mathbf{Q}^{\text{slow}}_{\text{i,i}} $ is the $i$th diagonal element of $\mathbf{Q}^{\text{slow}}$ and $\epsilon_{\text{CSP}}$ is a small threshold value.

Once the set of QSS species are selected, applying the QSS approximation results in a set of nonlinear algebraic equations for the concentrations of each QSS species, coupled with the remaining differential equations governing the non-QSS species.
Past efforts focused on solving this system of equations through iterative schemes, but convergence difficulties can arise due to deterioration of the QSS assumption when near and outside the validity range, leading to excessive computational cost~\cite{Law:2003wt}.
An alternative is to linearize the relations---assuming the coupling between QSS species is sparse in general---and generate an analytical solution for the concentrations of the QSS species.
Here, we present such a methodology, adopted from the approach established by Lu and Law~\cite{Lu:2006cn}, who also presented greater detail and explanation of this method.
We summarize the necessary steps here.

First, we must ensure the contribution of the nonlinear terms in the QSS equations is negligible such that these terms can be eliminated from the relations.
According to the QSS approximation, the net production rate of a QSS species is small compared to both the production and consumption rates.
This approximation results in a system of algebraic equations for the QSS species:
\begin{equation}
\label{E:qss}
\omega_{\text{P}, i} = \omega_{\text{C}, i} \quad i = 1, 2, \dots, N \;,
\end{equation}
with the species production and consumption rates expressed as
\begin{equation}
\omega_{\text{P}, i} = \sum_{j = 1}^{N_R} \nu_{i, j}^{\prime \prime} \Omega_j \quad \text{and} \quad
\omega_{\text{C}, i} = \sum_{j = 1}^{N_R} \nu_{i, j}^{\prime} \Omega_j \;,
\end{equation}
respectively, where $N$ is the number of QSS species and $N_R$ the number of irreversible reactions.
Note that unlike the previous reduction stages, this step requires all reactions to be irreversible.
The reaction rate is calculated using
\begin{equation}
\label{E:reac-rate}
\Omega_j = k_j \prod_{k = 1}^{N_S} x_{k}^{\nu_{k,j}^{\prime}} \;,
\end{equation}
where $k_j$ is the rate coefficient, $N_S$ the total number of species (both QSS and non-QSS), and $x_k$ the molar concentration of the $k$th species.
Equation~\eqref{E:qss} may be nonlinear due to the participation of multiple QSS species or a stoichiometric coefficient greater than one for a particular QSS species in a reaction.
However, due to the typically low concentration of QSS species (after an initial transient period), these nonlinear terms may not be important.
This importance can be quantified by calculating the normalized contribution of the nonlinear terms to the production and consumption rates of the $i$th QSS species, expressed as
\begin{align}
\pi_i &= \frac{ \sum_{j = 1}^{N_R} \nu_{i, j}^{\prime \prime} \Omega_j \delta_j }{ \omega_{\text{P}, i} } \;, \text{ and} \\
\kappa_i &= \frac{ \sum_{j = 1}^{N_R} \nu_{i, j}^{\prime} \Omega_j \delta_j }{ \omega_{\text{C}, i} } \;,
\end{align}
respectively, where
\begin{equation}
\label{E:delta-j}
\delta_j = \begin{cases}
1 & \text{if irreversible reaction \emph{j} involves} >1 \text{ QSS reactant,} \\
0 & \text{otherwise.} \end{cases}
\end{equation}

These measures of importance are similar to those used in DRG\slash DRGEP as well as unimportant reaction elimination, and here---as in those methods---the terms are considered unimportant if the values fall below a cutoff threshold.
The nonlinear terms may be neglected if
\begin{align}
\max_{k \in \lbrace \mathcal{D} \rbrace} \left( \max_{\text{all QSS species } i, k} \pi_i \right) <& \epsilon_{\text{nonlin}} \quad \text{and} \label{E:nonlin_prod} \\
\max_{k \in \lbrace \mathcal{D} \rbrace} \left( \max_{\text{all QSS species } i, k} \kappa_i \right) <& \epsilon_{\text{nonlin}} \;, \label{E:nonlin_con}
\end{align}
where $k$ is a reaction state, $\lbrace \mathcal{D} \rbrace$ the set of all reaction states of interest, and $\epsilon_{\text{nonlin}}$ a small user-defined threshold (e.g., \numrange{0.1}{0.2}).
If Eqs.~\eqref{E:nonlin_prod} and \eqref{E:nonlin_con} are satisfied using the criteria set by $\epsilon_{\text{nonlin}}$, all of the nonlinear contributions to QSS equations are deemed negligible and removed.
Note that $\epsilon_{\text{nonlin}}$ is used to evaluate the total contribution of all nonlinear terms rather than remove specific terms.

Once the nonlinear terms are eliminated, the QSS relations in Eq.~\eqref{E:qss} can be expressed using a system of linear equations, which Lu and Law~\cite{Lu:2006cn} termed the linearized QSS approximation (LQSSA):
\begin{equation}
C_i x_i = \sum_{k \neq i} P_{i k} x_k + P_{i 0} \quad i = 1, 2, \dots, N \label{E:lqssa} \;, \\
\end{equation}
where
\begin{align}
C_i &= \frac{ \omega_{\text{C} , i} }{ x_i } \;, \label{E:C_i} \\
P_{i k} &= \frac{ \sum_{j = 1}^{N_R} \nu_{i, j}^{\prime \prime} \Omega_j \sgn \left( \nu_{k, j}^{\prime} \right) }{ x_k } \;, \label{E:P_ik} \\
P_{i 0} &= \sum_{j = 1}^{N_R} \nu_{i, j}^{\prime \prime} \Omega_j \delta_j^{\prime} \;, \\
\sgn \left( \nu_{k, j}^{\prime} \right) &=
\begin{cases}
1 & \text{if } \nu_{k, j}^{\prime} > 0 \;, \\
0 & \text{if } \nu_{k, j}^{\prime} = 0 \;,
\end{cases} \quad \text{and}\\
\delta_j^{\prime} &= 
\begin{cases}
1 & \text{if irreversible reaction $j$ involves no QSS species as reactant,} \\
0 & \text{otherwise.} 
\end{cases}
\end{align}
Note that the consumption and production coefficients $C_i$, $P_{i k}$, and $P_{i 0}$ are independent of QSS species concentrations, and are either positive or zero.\footnote{Although the QSS species concentrations $x_i$ and $x_k$ appear in the denominators of Eqs.~\eqref{E:C_i} and \eqref{E:P_ik}, respectively, they are already present in the numerators through $\omega_{C,i}$ and $\Omega_j$ and therefore cancelled out.}
As with the initial nonlinear QSS relations, the system of equations given by Eq.~\eqref{E:lqssa} may be solved by iterative schemes, but could suffer the same computational difficulties.
In addition, the system could be solved through the typical Gaussian elimination, but its algorithmic complexity is a cubic function of $N$.
Instead, an analytic solution based on variable substitution and elimination offers an efficient approach for calculating the QSS species concentrations.
Now, the challenge becomes finding the best order for elimination by substitution that minimizes the number of operations required.
Lu and Law~\cite{Lu:2006cn} proposed using graph theory to identify the interdependence of QSS species.
We detail the construction of such a QSS graph (QSSG) in the following.

The system of LQSSA equations, as given by Eq.~\eqref{E:lqssa}, can be transformed to a form that offers a direct solution for each variable:
\begin{equation}
\label{E:qssg}
x_i = \sum_{k \neq i} A_{i k} x_k + A_{i 0} \quad i = 1, 2, \dots, N \;,
\end{equation}
where
\begin{equation}
A_{i k} = \frac{ P_{i k} }{ C_i } \quad \text{and} \quad A_{i 0} = \frac{ P_{i 0} }{C_i} \;.
\end{equation}
In this formulation, the solution for the concentration $x_i$ directly requires $x_k$ if $A_{i k} > 0$.
Similar to the concept used in DRG\slash DRGEP, the dependence of QSS species concentrations on one another can be mapped to a directed graph, where each QSS species is a graph node.
Edges between nodes exist when there is a direct dependence between species: the edge $x_i \rightarrow x_k$ exists if and only if $A_{i k} > 0$.
In some cases, Eq.~\eqref{E:qssg} may be explicit for all QSS species---meaning there is no interdependence---and the equations can be solved in the appropriate order without the need to substitute expressions and eliminate variables.
In general, though, the QSSG will consist of strongly coupled groups of QSS species that form cycles of dependence---these are known as the strongly connected components (SCCs) of the graph.
Intergroup coupling, on the other hand, is acyclic, such that an explicit elimination order of groups may be determined.

One important step is to prune the QSSG of unimportant edges, again in a similar manner to the elimination of unimportant species in the DRG and DRGEP methods.
While less important when a small number of QSS species exist, trimming the edges in a large graph ensures that the matrix \textbf{A} formed by the coefficients $A_{i k}$ is sparse, resulting in multiple groups rather than one large cyclic group made up of all the QSS species.
The importance of QSSG edges can be determined by calculating the normalized contribution of the $k$th QSS species to the production rate of the $i$th QSS species:
\begin{equation}
r_{i k} = \max_{ \lbrace \mathcal{D} \rbrace } \left( \frac{ A_{i k} x_k }{ \sum_{j \neq i} A_{i j} x_j + A_{i 0} } \right) \;,
\end{equation}
where $\max_{ \lbrace \mathcal{D} \rbrace }$ indicates taking the maximum value over the set of all reaction states of interest $\lbrace \mathcal{D} \rbrace$.
Unimportant QSSG edges are then identified and removed through comparison with a small cutoff threshold $\epsilon_{\text{QSS}}$, such that the remaining edges satisfy
\begin{equation}
x_i \rightarrow x_k \iff r_{i k} \geq \epsilon_{\text{QSS}} \;.
\end{equation}

After pruning the graph edges, the next step is to identify the SCCs and perform a topological sort~\cite{Cormen:2009uw}, which provides the order in which the SCCs are to be solved.
This is performed using the DIGRAPH\_ADJ\_COMPONENTS subroutine of Burkardt's GRAFPACK~\cite{Burkardt:2006}, with the algorithm originally taken from Thulasiraman and Swamy~\cite{Thulasiraman:1992}.
The adjacency matrix \textbf{E} of the QSSG, a necessary input, is formed by
\begin{equation}
E_{i k} = \begin{cases}
1 \;, \text{ if there is an edge } x_i \rightarrow x_k, \\
0 \;, \text{ otherwise.}
\end{cases}
\end{equation}
With the SCCs identified and sorted, the only remaining task is to solve for the intra-SCC species concentrations through variable elimination by substitution.
Lu and Law~\cite{Lu:2006cn} proposed a method to identify a near-optimal sequence for variable elimination, by calculating the normalized expansion cost $c_i$ of each variable $x_i$, defined as
\begin{equation}
\label{E:expcost}
\mathbf{c} = \mathbf{L} \mathbf{c} \;,
\end{equation}
where
\begin{align}
\mathbf{c} &= \left( c_1, c_2, \dots, c_M \right)^{\text{T}} \;, \\
L_{i k} &= \frac{ E_{i k} }{ \sum_{j = 1}^M E_{j k} } \;,
\end{align}
and $M$ is the number of QSS species in the current SCC.
Equation~\eqref{E:expcost} is an eigenvalue problem, where the column vector \textbf{c} is the eigenvector of \textbf{L} associated with the principal eigenvalue.
We solved this equation using the LAPACK subroutine DGEEV~\cite{Anderson:1999}, and selected the resulting eigenvector associated with the largest eigenvalue.
The values of \textbf{c} represent the relative expansion cost of each QSS species in the SCC, such that species with lower $c_i$ values should be eliminated from subsequent expressions first.
Therefore, species within the SCC are sorted in ascending order of $c_i$ for elimination by substitution.

Finally, we note that in some cases, the LQSSA---and therefore the analytic QSS solution---may not be valid when the contributions from the nonlinear terms ($\pi_i$ and $\kappa_i$) are not negligible.
For example, Lu and Law~\cite{Lu:2006cn} compared the terms' importance to detailed and skeletal mechanisms for ethylene, consisting of 70 and 33 species, respectively.
They found that while the terms were in fact small (between \numrange{0.1}{0.2}) for the skeletal mechanism, the same was not true for the detailed mechanism, where the nonlinear contributions to the production rates ($\pi_i$) were nearly unity in some cases.
It remains to be seen whether the LQSSA approximation is valid in the case of more complex skeletal mechanisms resulting from larger initial detailed mechanisms than ethylene.
When non-negligible nonlinear terms exist in the skeletal mechanism, two main options exist: (1) remove offending species from the QSS list, such that the nonlinear terms become negligible; and (2) develop a hybrid analytic-iterative solution scheme, where most of variables are calculated analytically and a small number of nonlinear terms are solved iteratively.
In the current study, no selected parameter values during the time-scale reduction stage resulted in such a situation, so we did not need to pursue either option.

\subsubsection{Reduction package}

\begin{figure}[tbp]
	\centering
	\includegraphics[width=0.9\linewidth]{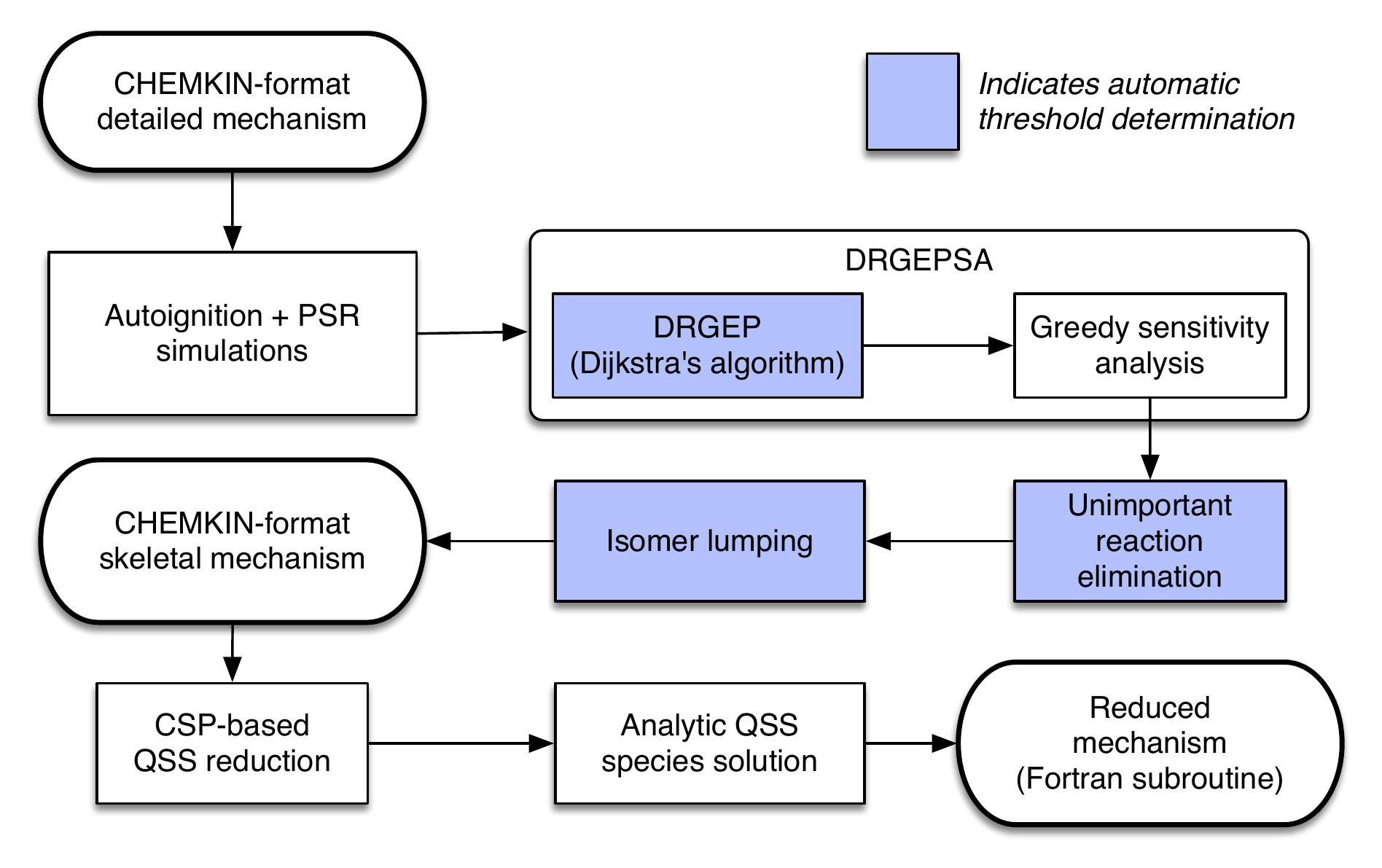}
	\caption{Flowchart depicting the various stages of the MARS reduction package. Stages highlighted in blue indicate those with automatic cutoff threshold determination based on the user-specified error limit.}
	\label{F:MARS-flowchart}
\end{figure}

The updated DRGEPSA method, the unimportant reaction elimination stage, the new isomer lumping stage, and the analytic QSS reduction stage were incorporated into the latest version of the Mechanism Automatic Reduction Software (MARS) package\cite{Niemeyer:2010bt,Niemeyer:2010,Niemeyer:2014}, as described by the flowchart in \ref{F:MARS-flowchart}.
This software is available by request to the authors.

%% MARS flowchart

A skeletal mechanism is generated as a collection of elementary reactions in the standard CHEMKIN format~\cite{Kee:1996vd}, which may be easily used with any chemical kinetics platform.
In contrast, a reduced mechanism requires a custom subroutine for evaluating the non-QSS species production rates along with the QSS species concentrations.
More implementation details involving reduced mechanisms may be found in, e.g., Chen~\cite{Chen:1997vq} or Sung et al.~\cite{Sung:1998gr,Sung:2001wa}.

%%%%%%%%%%%%%%%%%%%%%%%%%%%%%%%%%%%%%%%%%%%%%%%%%%%%%%%%%%%%%%%%%%%%%%%
\section{Results and Discussion}
%%%%%%%%%%%%%%%%%%%%%%%%%%%%%%%%%%%%%%%%%%%%%%%%%%%%%%%%%%%%%%%%%%%%%%%

\subsection{Mechanism reduction}

\subsubsection{HCCI-like conditions}

\begin{figure}[tbp]
	\centering
	\includegraphics[width=0.9\linewidth]{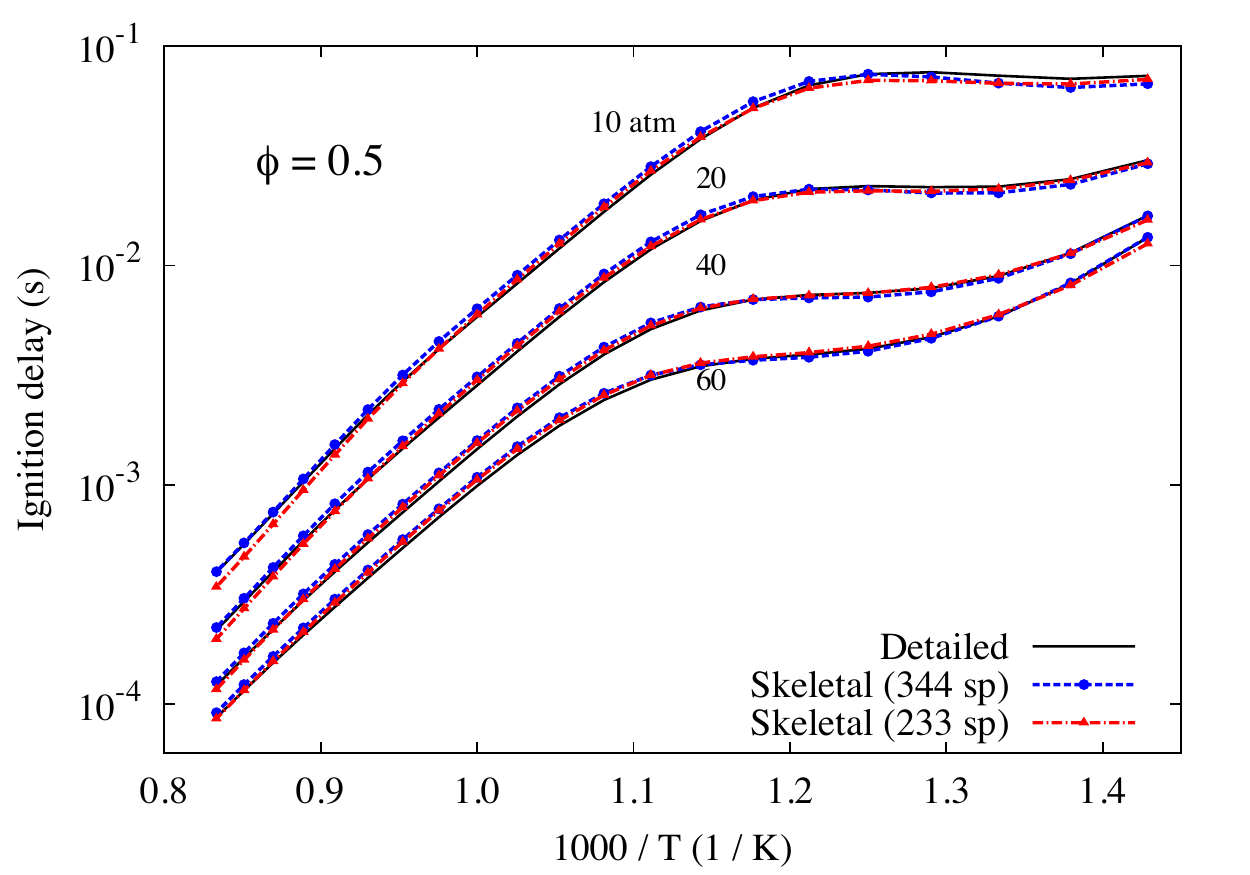}
	\caption{Autoignition validation of the DRGEPSA-produced skeletal mechanisms for the LLNL gasoline surrogate, corresponding to a \SI{10}{\percent} error limit, over a range of initial temperatures and pressures and at $\phi$ = 0.5 in air.
	The skeletal mechanisms with 344 and 233 species were generated using the initially informed and greedy sensitivity analysis algorithms, respectively.}
	\label{F:hcci-SA-mechs-phi-1.0}
\end{figure}

First, we compared the performance of the initially informed and greedy sensitivity analysis algorithms in DRGEPSA.
For the HCCI-like set of conditions, these algorithms resulted in skeletal mechanisms with 344 species and 1645 reactions, and 233 species and 1061 reactions, respectively, with maximum errors of \SI{9.3}{\percent} and \SI{8.9}{\percent} (based on the set of initial conditions given in \ref{T:conditions}).
In addition, we validated the ignition delay predictions of both skeletal mechanisms over the full target range of conditions; both mechanisms performed well, as shown in \ref{F:hcci-SA-mechs-phi-1.0} using $\phi$ = 0.5 for the LLNL gasoline surrogate in air.
The greater extent of reduction of the greedy sensitivity analysis---removing nearly 100 species more than the initially informed algorithm---while retaining a similar level of performance suggest the usefulness of the algorithm.

\begin{table}[tbp]
\begin{center}
\begin{tabular}{@{}l l l@{}}
\toprule
$\phi$ & $T$ (K) & $P$ (atm) \\
\midrule
1.0	&	800		& 	10 \\
1.0	&	750		&	60 \\
0.6	&	1200	&	60 \\
0.6	&	1100	&	10 \\
0.6	&	1000	&	60 \\
0.6	&	800		&	10 \\
0.6	&	750		&	10 \\
0.6	&	750		&	60 \\
0.2	&	800		&	60 \\
0.2	&	700		&	20 \\
0.2	&	800		&	20 \\
\bottomrule
\end{tabular}
\caption{Set of initial conditions used to generate skeletal mechanisms for the LLNL gasoline surrogate. Adopted from that used by Mehl et al.~\cite{Mehl:2011jn}.}
\label{T:conditions}
\end{center}
\end{table}

\begin{table}[tbp]
\begin{center}
\begin{tabular}{@{}l l l l@{}}
\toprule
Stage 	& \# Species	& \# Reactions	& Max.\ error	\\
\midrule
Detailed & 1388 	& 5933		& \\
DRGEP	& 471		& 2434		& \SI{3.8}{\percent} \\
(Greedy) SA	& 233		& 1061		& \SI{8.9}{\percent} \\ 
Reac.\ elim	& 233	& 910		& \SI{8.3}{\percent} \\
Isomer lump & 213	& 910		& \SI{9.5}{\percent} \\
CSP\slash QSS	& 148 (65 QSS)	& 1809 & \SI{9.3}{\percent} \\
\bottomrule
\end{tabular}
\caption{Summary of results from mechanism reduction stages for HCCI-like target range of conditions.
The reaction number discrepancy between the isomer lumping and CSP\slash QSS stage resulted from the need to convert reversible reactions into two irreversible reactions.}
\label{T:hcci-results}
\end{center}
\end{table}

%% HCCI greedy vs initially informed SA mechanisms

Using the greedy SA-generated mechanism for the remaining analysis, the results from all reduction stages are summarized in \ref{T:hcci-results}.
For DRGEPSA, the cutoff threshold and upper threshold for sensitivity analysis were \num{7e-3} and \num{0.1}, respectively.
The unimportant reaction elimination algorithm iteratively selected a cutoff threshold for reactions of \num{6e-3}.
Out of 40 potential isomer groups containing a total of 125 species, only seven groups with 27 isomers were selected using an error-based coefficient of determination cutoff of \num{0.995}.
The isomers and their groups are listed in \ref{T:hcci-isomers}.
The small number of isomers selected for lumping suggested that isomer mass fractions are less correlated at the lower temperatures experienced in this range of conditions; Luo et al.~\cite{Luo:2012cr} experienced similar behavior when generating a skeletal mechanism for biodiesel surrogates valid at low temperatures.
For the CSP-based QSS reduction stage, we selected $\alpha_{\text{CSP}}$ = 100, $\epsilon_{\text{CSP}}$ = \num{1e-4}, and $\epsilon_{\text{nonlin}}$ = 0.1, using the QSSG cutoff threshold $\epsilon_{\text{QSS}}$ = 0.01.

\begin{table}[tbp]
\renewcommand{\arraystretch}{1.25}
\begin{center}
\begin{tabular}{@{}l l@{}}
\toprule
Group 	& Isomers	\\ \midrule
\multirow{2}{*}{\ce{C7H14OOH$\hyphen$O2}} & \ce{C7H14OOH$1\hyphen3$O2}, \ce{C7H14OOH$2\hyphen4$O2}, \\ 
 & \ce{C7H14OOH$3\hyphen5$O2}, \ce{C7H14OOH$4\hyphen2$O2} \\
\multirow{2}{*}{\ce{NC7KET}} & \ce{NC7KET$13$}, \ce{NC7KET$24$}, \ce{NC7KET$32$}, \\
 & \ce{NC7KET$35$}, \ce{NC7KET$42$} \\
\multirow{2}{*}{\ce{C8H16OOH}} & \ce{AC8H16OOH$\hyphen$B}, \ce{AC8H16OOH$\hyphen$C}, \\
 & \ce{BC8H16OOH$\hyphen$A}, \ce{DC8H16OOH$\hyphen$B} \\
\ce{C8H16OOH$\hyphen$O2} & \ce{AC8H16OOH$\hyphen$BO2}, \ce{BC8H16OOH$\hyphen$AO2}, \ce{DC8H16OOH$\hyphen$BO2} \\
\ce{IC8KET} & \ce{IC8KETAB}, \ce{IC8KETBA}, \ce{IC8KETBD}, \ce{IC8KETDB} \\
\ce{C7H15O2} & \ce{C7H15O2$\hyphen1$}, \ce{C7H15O2$\hyphen2$}, \ce{C7H15O2$\hyphen3$}, \ce{C7H15O2$\hyphen4$} \\
\ce{C8H17O2} & \ce{AC8H17O2}, \ce{BC8H17O2}, \ce{DC8H17O2} \\
\bottomrule
\end{tabular}
\caption{Lumped isomer groups selected for HCCI-like mechanism reduction.
Refer to Mehl et al.~\cite{Mehl:2011cn,Mehl:2011jn} for the species nomenclature.}
\label{T:hcci-isomers}
\end{center}
\end{table}

\begin{figure}[tbp]
\begin{center}
\begin{subfigure}[b]{0.45\textwidth}
	\begin{center}
	\includegraphics[width=\textwidth]{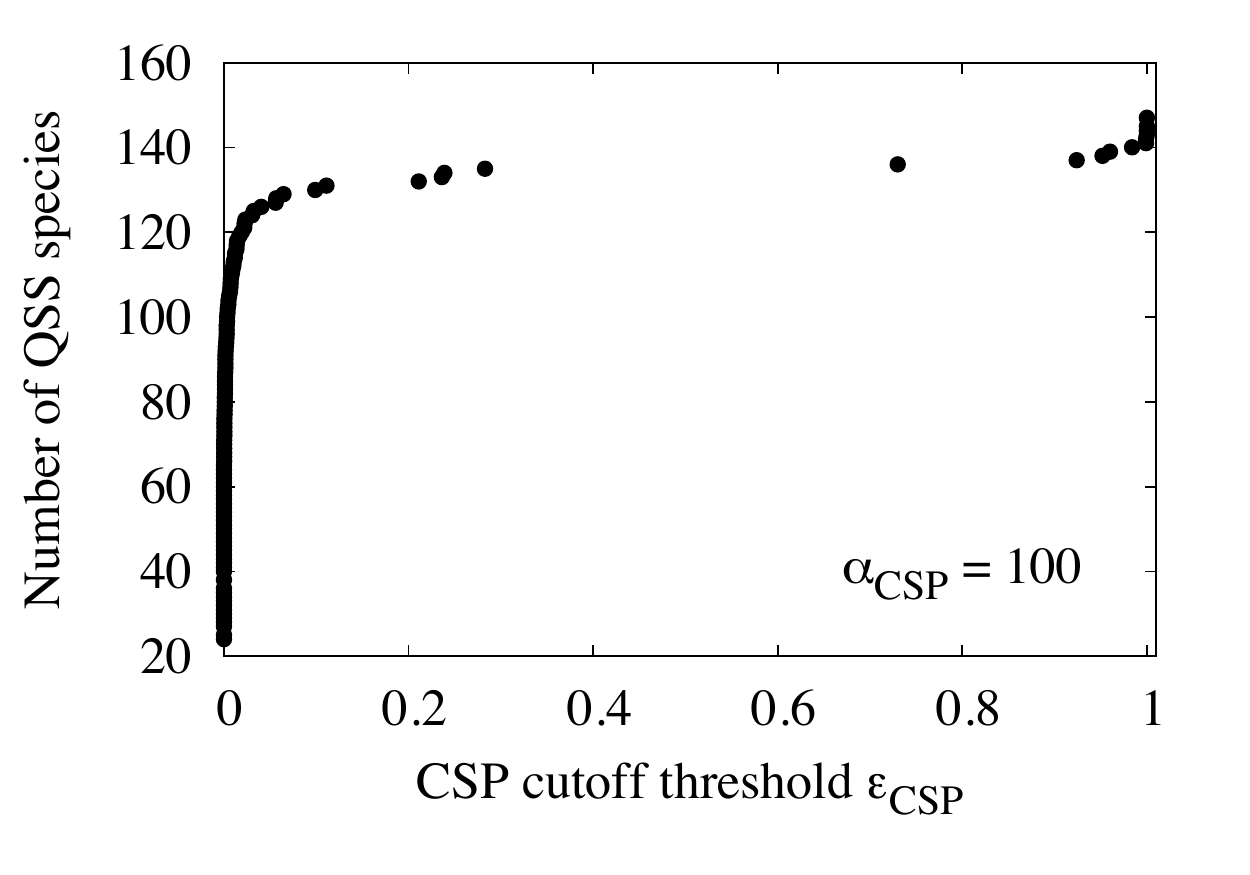}
	\caption{Linear plot}
	\label{F:hcci-qss-threshold-linear}
	\end{center}
\end{subfigure}
\begin{subfigure}[b]{0.45\textwidth}
	\begin{center}
	\includegraphics[width=\textwidth]{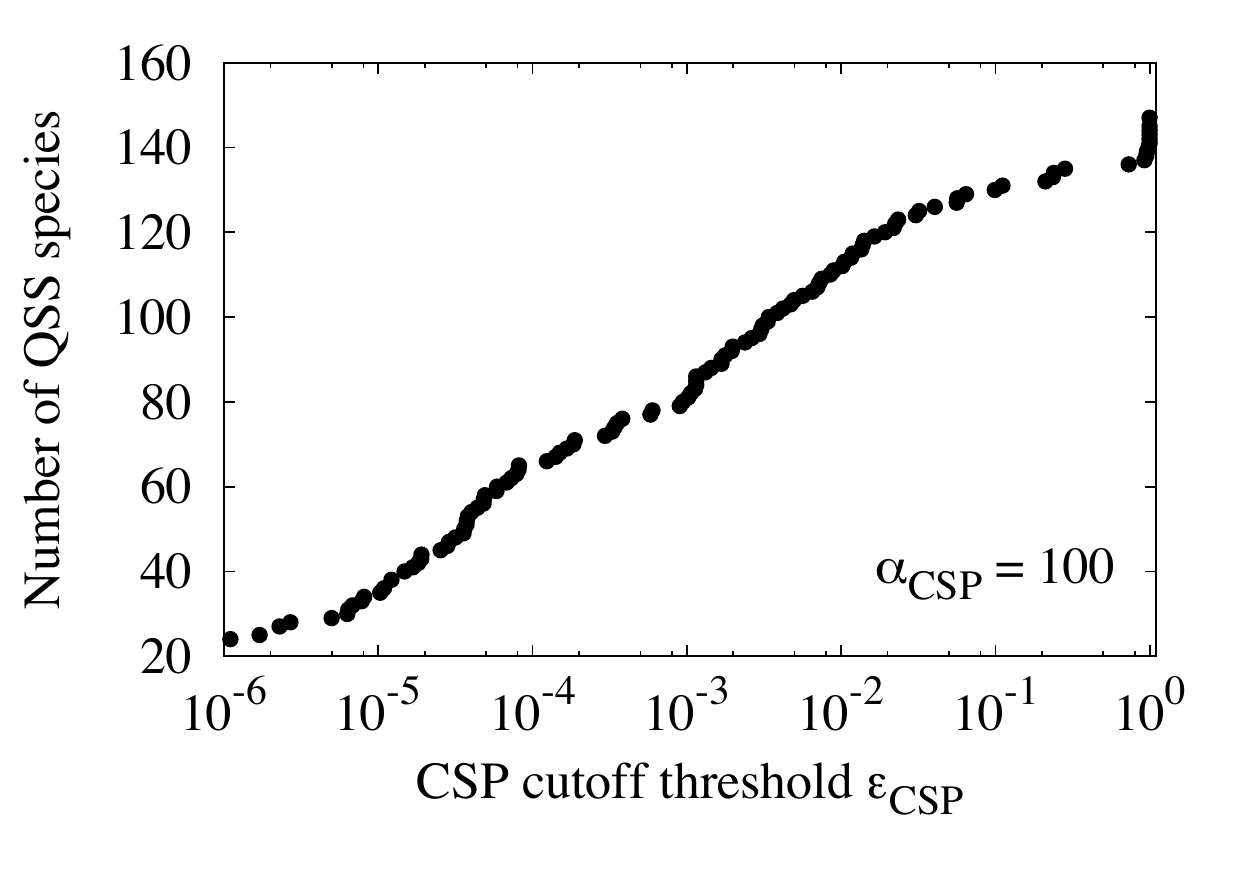}
	\caption{Semilogarithmic plot}
	\label{F:hcci-qss-threshold-log}
	\end{center}
\end{subfigure}
\caption{Number of QSS species as a function of CSP cutoff threshold $\epsilon_{\text{CSP}}$ for HCCI-like range of conditions. Gaps in points indicate no change in number of QSS species for increasing cutoff threshold value.}
\label{F:hcci-qss-threshold}
\end{center}
\end{figure}

We note that unlike the skeletal reduction parameters ($\epsilon_{\text{EP}}$, $\epsilon_{\text{reac}}$, $\epsilon_{\text{isom}}$) determined automatically based on error limit, the QSS reduction parameters ($\alpha_{\text{CSP}}$, $\epsilon_{\text{CSP}}$, and $\epsilon_{\text{QSS}}$) were chosen based on trial and error.
Lu and Law~\cite{Lu:2008dp,Lu:2008bi} used ``jumps'' in the numbers of QSS species as a function of $\epsilon_{\text{CSP}}$ to select the best values for methane and \emph{n}-heptane reduced mechanisms, choosing 0.1 in both cases.
\ref{F:hcci-qss-threshold-linear} shows this relationship for the current situation plotted in a linear format, for the skeletal mechanism following isomer lumping with 213 species and 910 reactions.
Similar jumps in number are observed---in fact, \ref{F:hcci-qss-threshold-linear} appears nearly identical to those shown by Lu and Law~\cite{Lu:2008dp,Lu:2008bi}---but we found that more care must be taken in selecting $\epsilon_{\text{CSP}}$.
For example, by taking the value 0.15 based on the first major jump in \ref{F:hcci-qss-threshold-linear}, we produced a reduced mechanism with 82 non-QSS and 131 QSS species.
This reduced mechanism performed poorly, demonstrating a maximum error of \SI{825.6}{\percent}.
In addition, the maximum contribution of the (removed) nonlinear QSS terms $\pi_i$ and $\kappa_i$ were both non-negligible, around 0.6.
Practically, the large number of QSS species resulted in an incredibly complex analytic solution for the QSS species concentrations, producing a Fortran source code file with over \num{60000} lines.

Obviously, $\epsilon_{\text{CSP}}$ = 0.15 was too large for this case.
In order to select a more appropriate value, we replotted the relationship between $\epsilon_{\text{CSP}}$ and number of QSS species into a semilogarithmic format, as shown in \ref{F:hcci-qss-threshold-log}.
Viewed in this manner, a number of jumps were observed.
We selected \num{1e-4} based on the location of one jump (as well as trial-and-error), and this resulted in a reduced mechanism with 148 non-QSS and 65 QSS species that performed within the \SI{10}{\percent} error limit; additionally, the nonlinear contributions remained negligible ($<$ \num{5e-6}).
A more systematic approach to determining the optimal $\epsilon_{\text{CSP}}$ values is warranted, which will be pursued in future work.
For example, one option may be to select $\epsilon_{\text{CSP}}$ based on the values of the contribution of nonlinear terms.

Finally, the parameter $\epsilon_{\text{QSS}}$ was chosen as 0.01 in order to simplify the QSS analytic solution without introducing significant error.
Using $\epsilon_{\text{QSS}} = 0$ (i.e., retaining the entire QSSG) produced a reduced mechanism that performed with a maximum error of \SI{9.3}{\percent}.
However, this resulted in a Fortran source code file unable to be compiled by some compilers due to long expressions: the longest expression required over 1300 lines with about 200 characters per line, over \num{260000} total characters in a single statement.
Raising $\epsilon_{\text{QSS}}$ to 0.01 simplified the system considerably, such that the longest expression only needed 13 lines, without increasing error.

%%%%%%%%%%%%%%%%%%%%%%%%%%%%%%%%%%%%%%%%%%%%%%%%%%%
\subsubsection{SI\slash CI conditions}

\begin{figure}[tbp]
	\centering
	\includegraphics[width=0.9\linewidth]{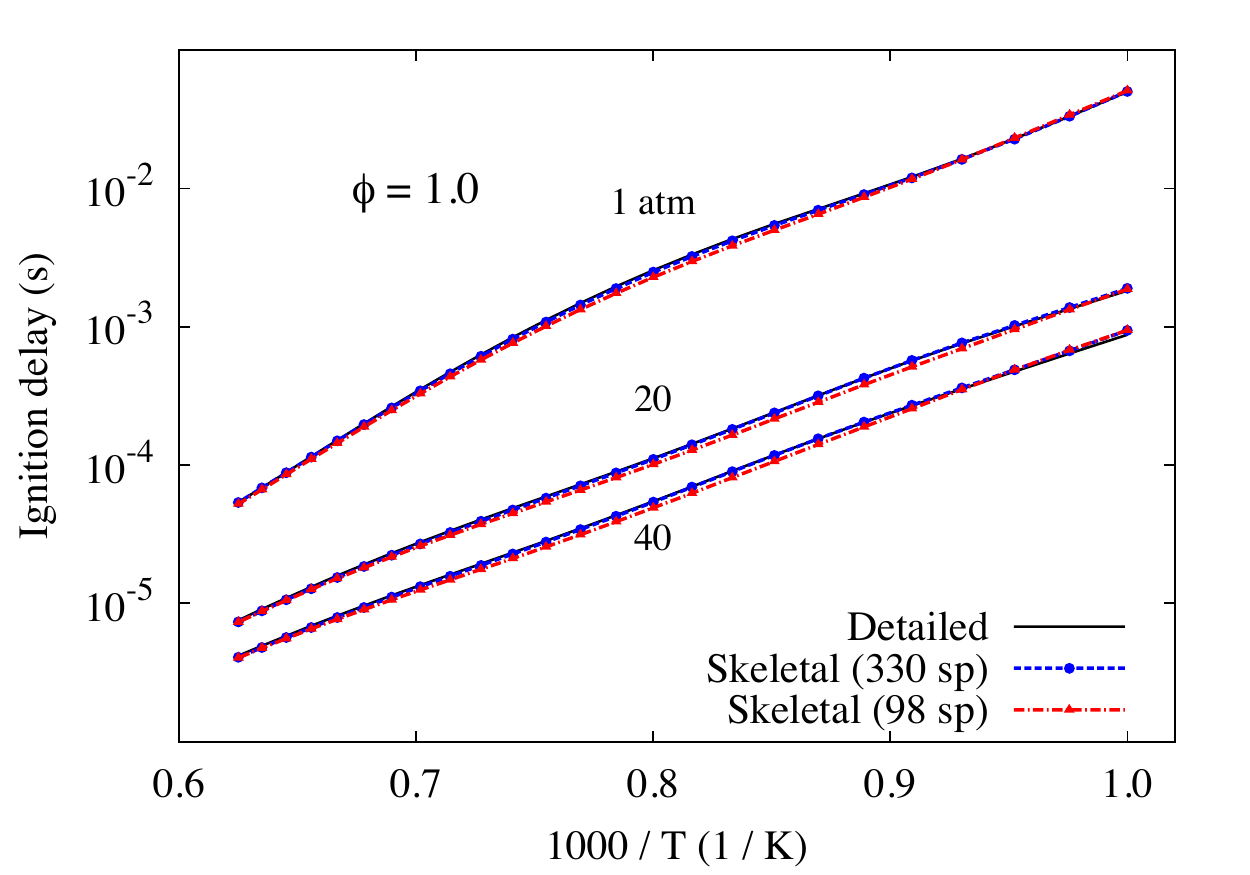}
	\caption{Autoignition validation of the DRGEPSA-produced skeletal mechanisms generated using a \SI{10}{\percent} error limit for the LLNL gasoline surrogate in air at high temperatures relevant to SI\slash CI conditions, over a range of initial temperatures and pressures and at $\phi$ = 1.0.
	The skeletal mechanisms with 330 and 98 species were generated using the initially informed and greedy sensitivity analysis algorithms, respectively.}
	\label{F:high-temp-SA-mechs-phi-1.0}
\end{figure}

As in the previous section, we first compared the results of the initially informed and greedy sensitivity analysis algorithms in order to further investigate the new algorithm.
Using both autoignition and PSR simulation data with $\epsilon^* = 0.5$, the initially informed algorithm generated a skeletal mechanism with 330 species and 1951 reactions, with a maximum error of \SI{8.4}{\percent}.
In contrast, the greedy algorithm gave a mechanism with 99 species and 611 reactions, predicting ignition delay with a maximum error of \SI{9.1}{\percent} over the set of initial conditions.
\ref{F:high-temp-SA-mechs-phi-1.0} shows the ignition delay predictions of the skeletal mechanisms produced by the initially informed and greedy algorithms for a range of initial temperatures and pressures at $\phi = 1.0$; both reproduced the calculations of the detailed mechanism at all conditions nearly identically, although clearly the greedy algorithm produced a notably smaller skeletal mechanism.

%% high-temp greedy vs initially informed SA mechanisms

\begin{table}[tbp]
\begin{center}
\begin{tabular}{@{}l l l l@{}}
\toprule
SA Algorithm 	& $\epsilon^*$ & PSR & \# Species \\
\midrule
Initially informed & 0.5 & yes & 330 \\
Greedy & 0.5 & yes & 99 \\
Initially informed & 0.1 & no & 249 \\
Greedy & 0.1 & no & 137 \\ % bad PSR/FS
Initially informed & 0.01 & no & 315 \\
Greedy & 0.01 & no & 314 \\
\bottomrule
\end{tabular}
\caption{Comparison of various reduction parameters and results for SI\slash CI reduction, where ``PSR'' indicates the use of PSR data in the reduction.
All skeletal reductions used a \SI{10}{\percent} error limit.}
\label{T:high-temp-comparison}
\end{center}
\end{table}

In order to study the effects of varying $\epsilon^*$ and using PSR in the reduction procedure on the two sensitivity analysis algorithms, we also generated additional skeletal mechanisms using different combinations of parameters as summarized in \ref{T:high-temp-comparison}.
Interestingly, the low and high extremes in terms of skeletal mechanism sizes resulted from the greedy and initially informed SA algorithms, respectively, using autoignition and PSR data with a relatively high $\epsilon^*$ value.
The greedy algorithm combined with a typical $\epsilon^*$ value (e.g., 0.1) without including PSR data resulted in a skeletal mechanism that not only contained a larger number of species but also performed poorly outside of autoignition (e.g., PSR, laminar flame speed calculations); this occurred due to the overaggressive removal of species unimportant to autoignition but important in other phenomena.
Niemeyer and Sung~\cite{Niemeyer:2014} previously demonstrated this issue when reducing detailed mechanisms for multicomponent surrogates, and suggested decreasing $\epsilon^*$ would prevent the removal of these species.
Unfortunately, as shown here, the cost of this solution is a larger resulting mechanism size.
Interestingly, with $\epsilon^* = 0.01$ and without using PSR data both algorithms produced similarly sized skeletal mechanisms; due to the severe limitations on the number of limbo species, the improved greedy algorithm was not able to perform more effectively.
Overall, our results suggest that using the greedy SA algorithm combined with PSR data (in addition to the default autoignition) and a high $\epsilon^*$ value is the best approach in terms of both the size and performance of the resulting skeletal mechanism.
Therefore, the skeletal mechanism associated with that approach is used for the following results and discussion.

\begin{table}[tbp]
\begin{center}
\begin{tabular}{@{}l l l l@{}}
\toprule
Stage 	& \# Species	& \# Reactions	& Max.\ error	\\
\midrule
Detailed & 1388 	& 5933		& \\
DRGEP	& 415		& 2362		& \SI{8.4}{\percent} \\
(Greedy) SA	& 99		& 611		& \SI{9.1}{\percent} \\ 
Reac.\ elim	& 99	& 513		& \SI{9.3}{\percent} \\
Isomer lump & 97	& 512		& \SI{9.5}{\percent} \\
CSP\slash QSS	& 79 (18 QSS)	& 1018 & \SI{11.5}{\percent} \\
\bottomrule
\end{tabular}
\caption{Summary of results from mechanism reduction stages for SI\slash CI-engine target range of conditions.
The reaction number discrepancy between the isomer lumping and CSP\slash QSS stage resulted from the need to convert reversible reactions into two irreversible reactions.}
\label{T:hightemp-results}
\end{center}
\end{table}

\ref{T:hightemp-results} summarizes the results of each reduction stage.
The DRGEP algorithm automatically selected a cutoff threshold of \num{3e-3}, and as mentioned above we set $\epsilon^* = 0.5$ for the (greedy) sensitivity analysis upper threshold.
The unimportant reaction elimination algorithm iteratively selected a cutoff threshold for reactions of \num{4e-3}.
Out of eight potential isomer groups containing a total of 21 species, the isomer lumping algorithm selected only one group with three isomers (\ce{C7H15$\hyphen2$}, \ce{C7H15$\hyphen3$}, and \ce{C7H15$\hyphen4$}) using an automatically selected coefficient of determination cutoff of \num{0.994}.
For the CSP-based QSS reduction stage, based on the HCCI reduction results we selected $\alpha_{\text{CSP}}$ = 100, $\epsilon_{\text{CSP}}$ = \num{1e-4}, and $\epsilon_{\text{nonlin}}$ = 0.1, using the QSSG cutoff threshold $\epsilon_{\text{QSS}}$ = 0.01.
Using these combinations of parameters, none of the QSS equations contained nonlinear terms and so no approximation was required here.

\begin{figure}[tbp]
\begin{center}
\begin{subfigure}[b]{0.45\textwidth}
	\begin{center}
	\includegraphics[width=\textwidth]{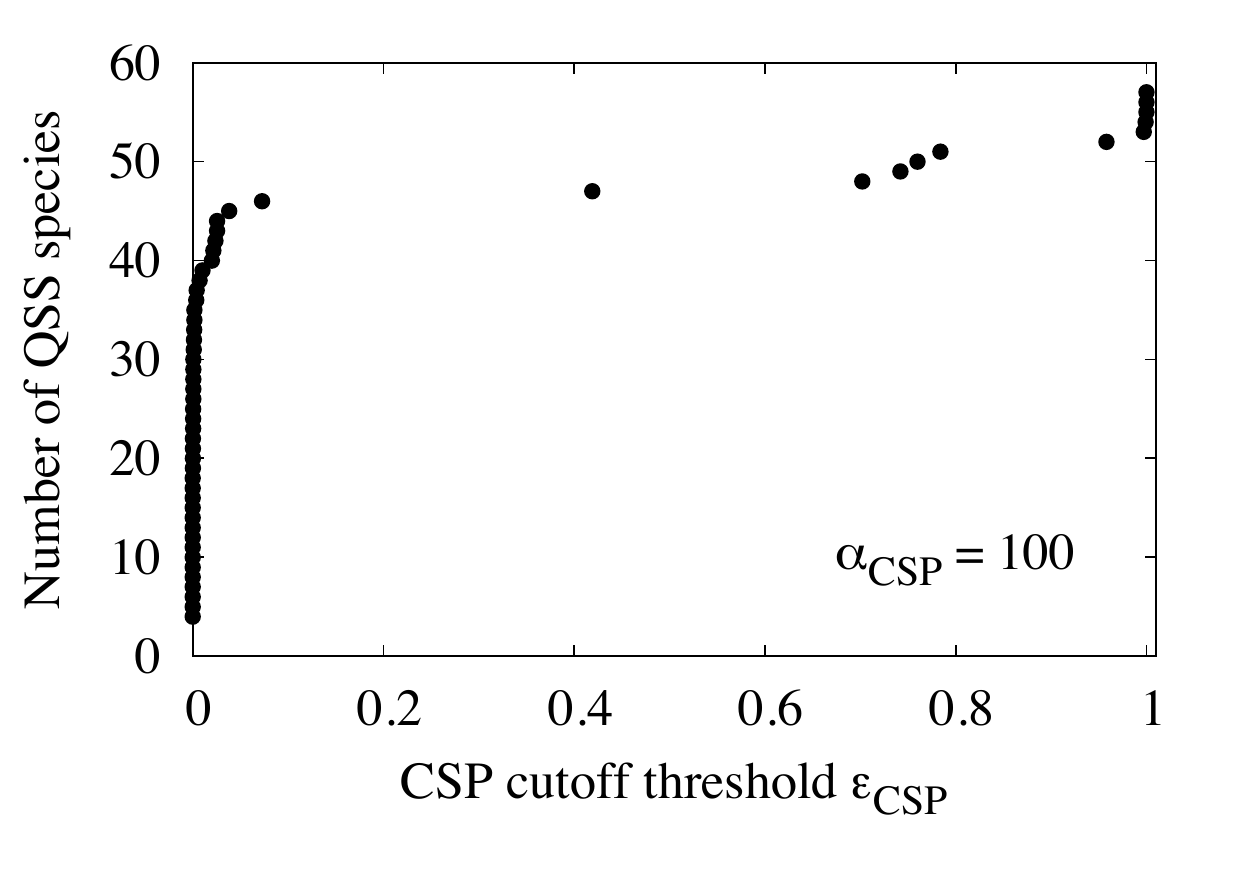}
	\caption{Linear plot}
	\label{F:si-qss-threshold-linear}
	\end{center}
\end{subfigure}
\begin{subfigure}[b]{0.45\textwidth}
	\begin{center}
	\includegraphics[width=\textwidth]{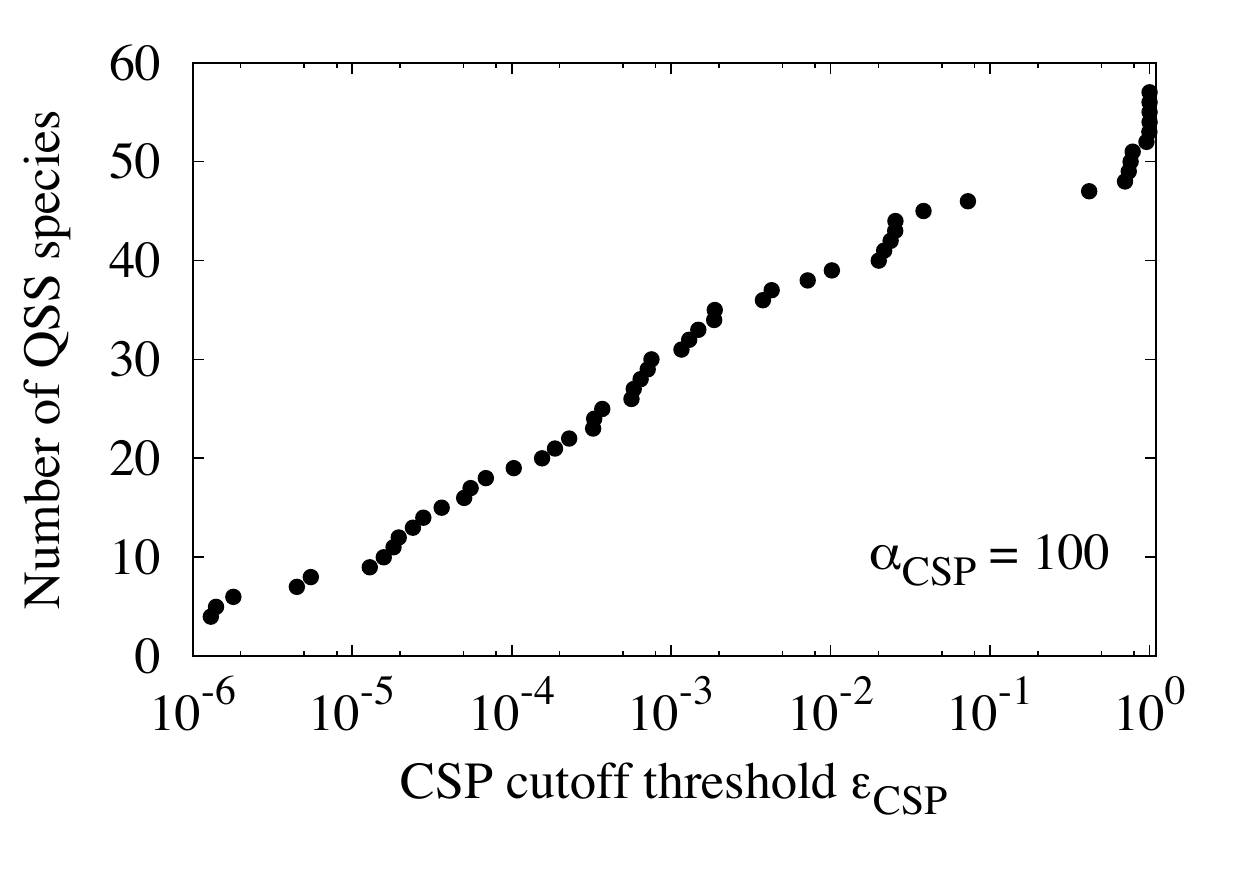}
	\caption{Semilogarithmic plot}
	\label{F:si-qss-threshold-log}
	\end{center}
\end{subfigure}
\caption{Number of QSS species as a function of CSP cutoff threshold $\epsilon_{\text{CSP}}$ for the SI\slash CI-like range of conditions. Gaps in points indicate no change in number of QSS species for increasing cutoff threshold value.}
\label{F:si-qss-threshold}
\end{center}
\end{figure}

Regarding the CSP parameters, \ref{F:si-qss-threshold} shows the relationship between the number of QSS species and the CSP cutoff threshold $\epsilon_{\text{CSP}}$; the trends resemble those seen in \ref{F:hcci-qss-threshold} for the HCCI-condition reduction, although even more dramatic jumps are observed (e.g., near $\epsilon_{\text{CSP}} =$ \numlist{0.1;0.4;0.8}).
In this case, selecting $\epsilon_{\text{CSP}} = 0.1$ led to a reduced mechanism with 51 non-QSS and 46 QSS species that performed well in both the autoignition and PSR calculations considered; however, convergence problems arose during laminar flame speed calculations.
Therefore, the lower value of \num{1e-4} was selected as described above, resulting in 79 non-QSS and 18 QSS species.

Finally, we note that in \ref{T:hightemp-results} the final reduced mechanism with 79 non-QSS species slightly exceeded the \SI{10}{\percent} error limit set for use in the automated skeletal reduction stages.
The error above the limit only appears for autoignition near \SI{1000}{\kelvin} at high pressure; the error for other autoignition conditions and all PSR conditions falls below the \SI{10}{\percent} limit.
This violation of the set error limit occurs due to the manual adjustment of the CSP\slash QSS parameters, and motivates future work into more automated determination of the time-scale reduction parameters.

\subsection{Mechanism validation}

\subsubsection{HCCI-like conditions}

\begin{figure}[tbp]
	\centering
	\includegraphics[width=0.75\linewidth]{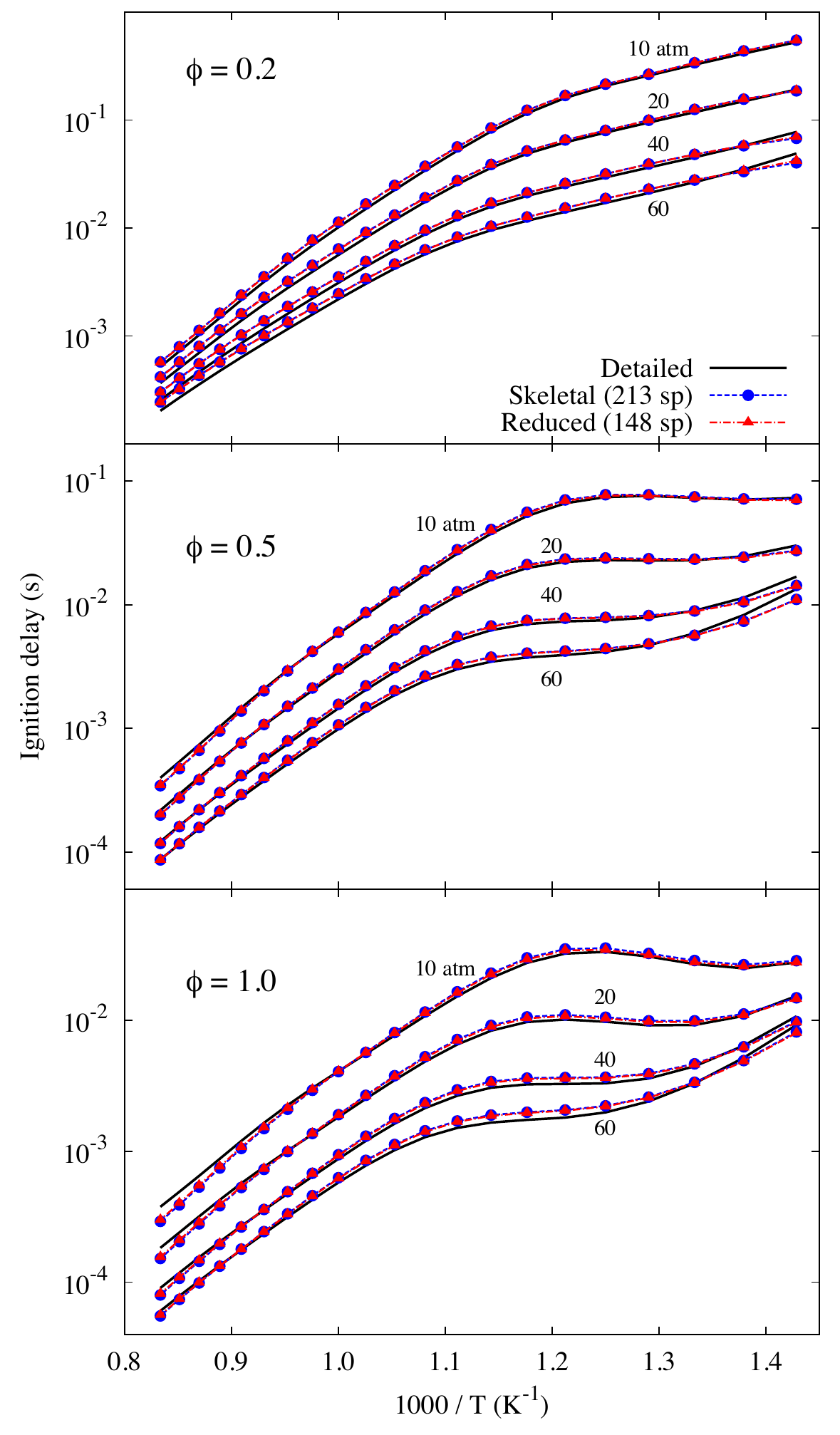}
	\caption{Autoignition validation of the skeletal (213 species) and reduced (148 species) mechanisms for the LLNL gasoline surrogate targeted at HCCI-like conditions.
	Ignition delays were calculated over a range of initial temperatures, pressures, and equivalence ratios.}
	\label{F:hcci-ign-delay}
\end{figure}

We performed validation of the skeletal and reduced mechanisms for the HCCI-like condition set using constant-volume autoignition simulations over a range of initial conditions for temperature, pressure, and equivalence ratio.
\ref{F:hcci-ign-delay} shows the comparison of the ignition delays calculated by the detailed mechanism, skeletal mechanism resulting from the isomer lumping stage, and final reduced mechanism.
Both the skeletal and reduced mechanisms performed well over the full range of conditions with the maximum error within \SI{10}{\percent}, despite some visible discrepancies at high temperatures and low temperatures\slash high pressures.
Both mechanisms captured the negative temperature coefficient regimes for $\phi$ = \numlist{0.5;1.0}.

%% HCCI skeletal and reduced mechanisms

\begin{figure}[tbp]
	\centering
	\includegraphics[width=0.75\linewidth]{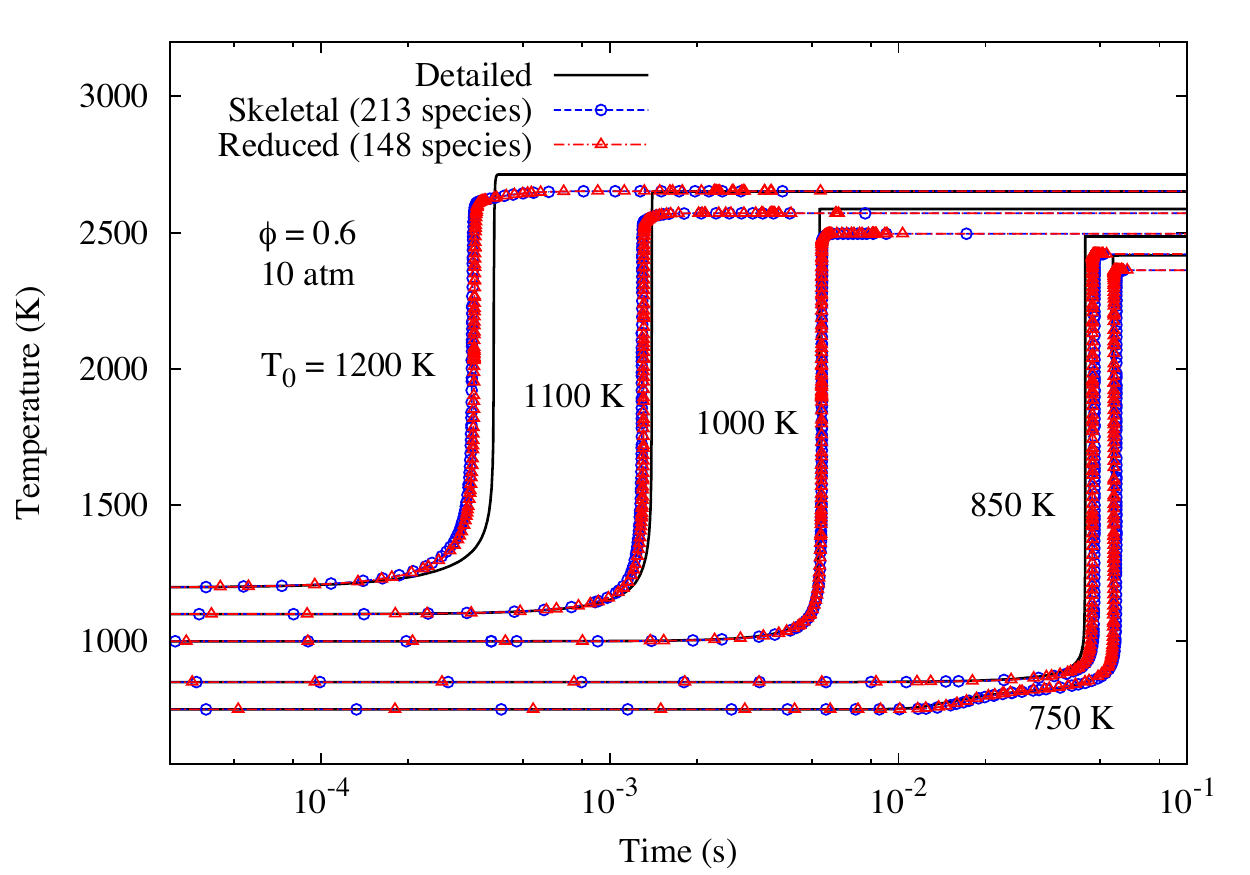}
	\caption{Comparison of temperature profiles for various initial temperatures at \SI{10}{\atm} and $\phi$ = 0.6 calculated by the detailed mechanism and skeletal (213 species) and reduced (148 species) mechanisms targeted at HCCI-like conditions.}
	\label{F:hcci-temp-10atm}
\end{figure}

\begin{figure}[tbp]
	\centering
	\includegraphics[width=0.75\linewidth]{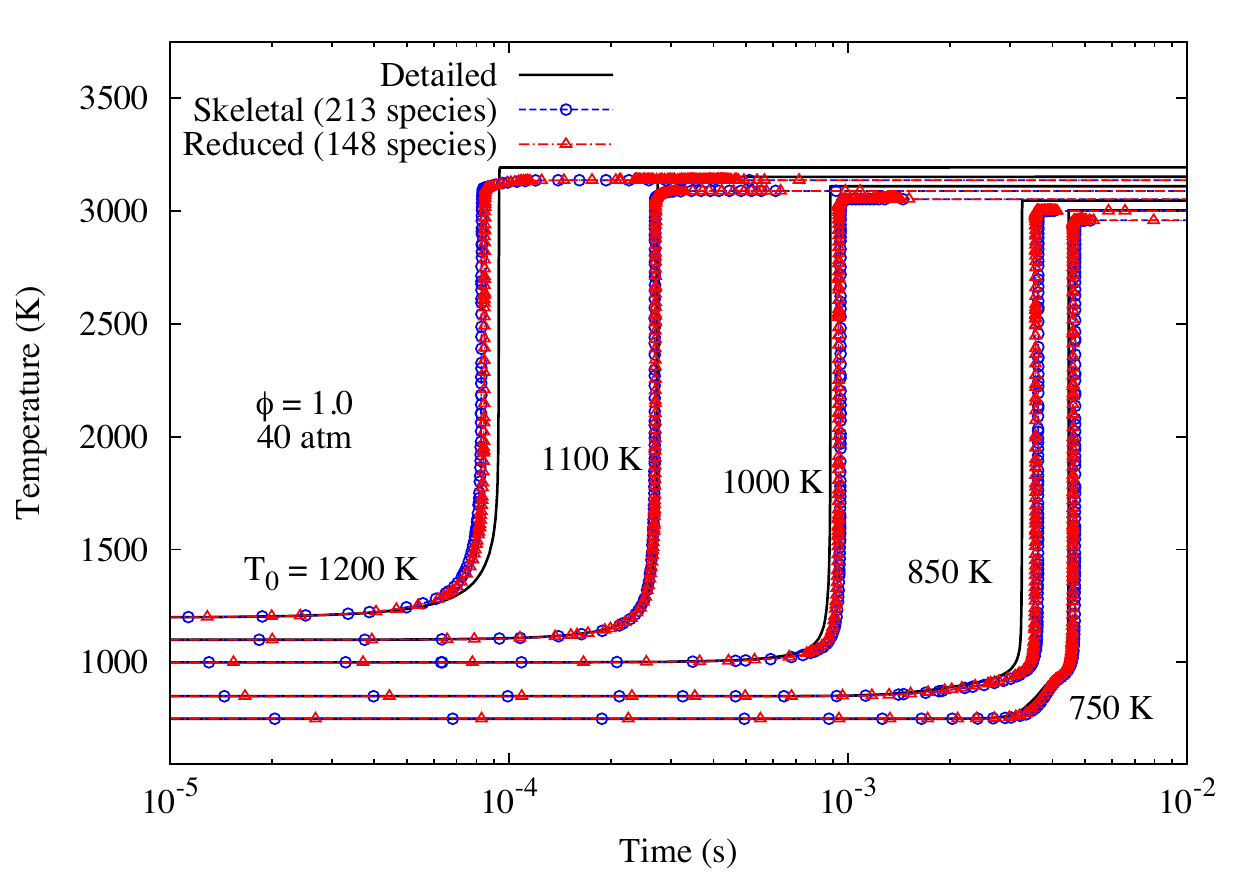}
	\caption{Comparison of temperature profiles for various initial temperatures at \SI{40}{\atm} and $\phi$ = 1.0 calculated by the detailed mechanism and skeletal (213 species) and reduced (148 species) mechanisms targeted at HCCI-like conditions.}
	\label{F:hcci-temp-40atm}
\end{figure}

\begin{figure}[tbp]
	\centering
	\includegraphics[width=0.75\linewidth]{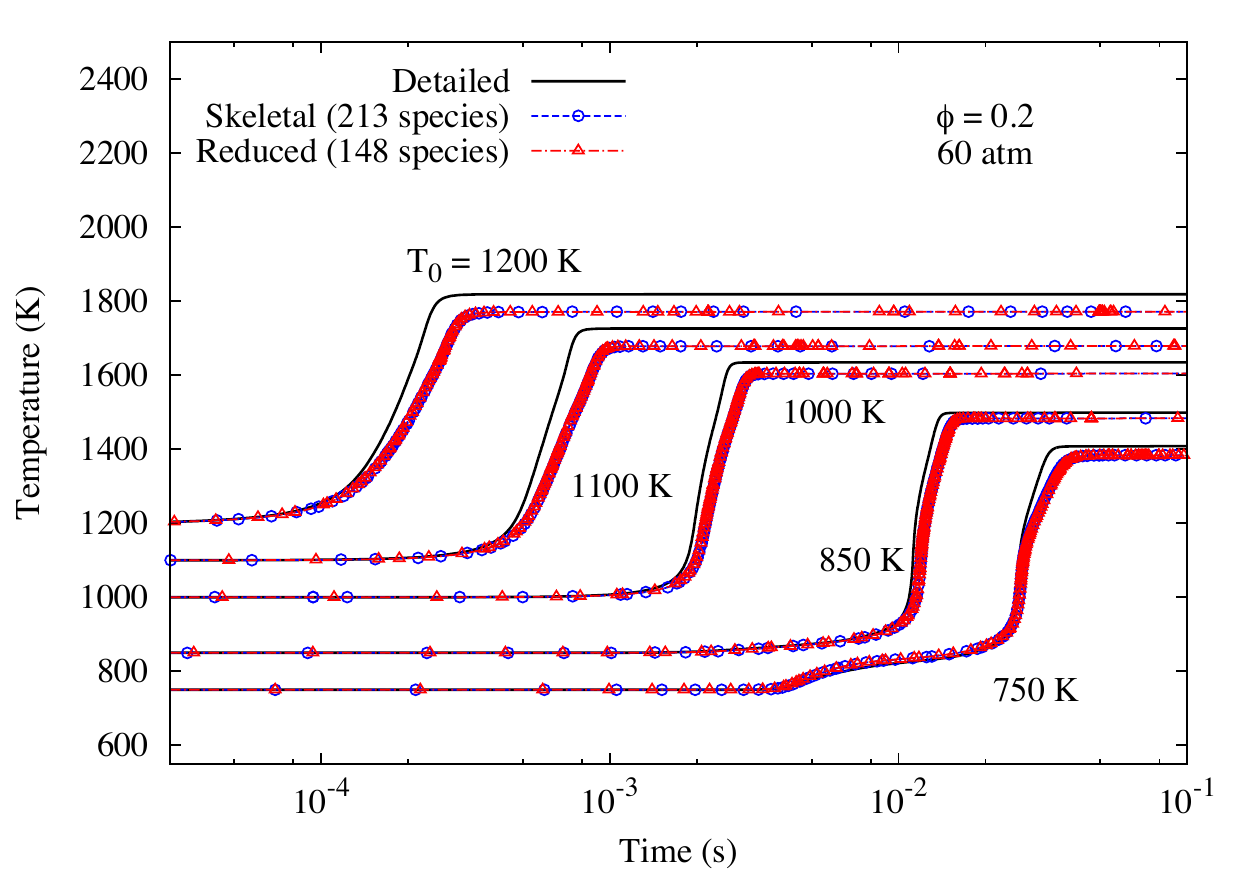}
	\caption{Comparison of temperature profiles for various initial temperatures at \SI{60}{\atm} and $\phi$ = 0.2 calculated by the detailed mechanism and skeletal (213 species) and reduced (148 species) mechanisms targeted at HCCI-like conditions.}
	\label{F:hcci-temp-60atm}
\end{figure}

Encouraged by the accurate prediction of ignition delay shown by the skeletal and reduced mechanisms, we also performed a more detailed validation analysis by comparing the temperature profiles for constant-volume autoignition initiating at different temperatures.
As we showed previously~\cite{Niemeyer:2014}, predicting the point of ignition does not guarantee prediction of time- or spatially-varying scalars.
\ref{F:hcci-temp-10atm}, \ref{F:hcci-temp-40atm}, and \ref{F:hcci-temp-60atm} show comparisons of temperature profiles for various initial temperatures at \SI{10}{\atm} and $\phi$ = 0.6, \SI{40}{\atm} and $\phi$ = 1.0,  and \SI{60}{\atm} and $\phi$ = 0.2, respectively.
In general, both the skeletal and reduced mechanisms accurately reproduced the temperature profiles.
Notably, the skeletal and reduced mechanisms performed indistinguishably.

%% HCCI skeletal and reduced mechanisms

Slight discrepancies in the point of ignition within the specified error limit were observed, and for certain conditions the skeletal and reduced mechanisms also underpredicted the post-ignition temperature slightly.
Luo et al.~\cite{Luo:2012cr} observed similar errors for some conditions in their biodiesel skeletal mechanism produced using DRGASA with autoignition, PSR, and jet stirred reactor data.
Comparing post-ignition temperatures calculated by skeletal mechanisms at varying stages in the reduction procedure, we traced this error back to the sensitivity analysis stage---the DRGEP-generated mechanism closely matched the detailed mechanism.
Both the initially informed and greedy algorithms introduced this error.
Underprediction of the post-ignition temperature is not caused by an error in equilibrium temperature; the values calculated by the skeletal mechanisms match those predicted by the detailed mechanism exactly.
Instead, our analysis showed that this error is caused by the removal of species during the sensitivity analysis stage.
In particular, the elimination of the reactions 
\begin{align*}
\cee{ACC6H10 + HO2 &<-> ACC6H9-D + H2O2 \text{ and} \tag{R1}\\
     YC7H14 + OH &<-> YC7H13-Y$2$ + H2O \text{,} \tag{R2}}
\end{align*}
through the removal of the species \ce{ACC6H9-D} and \ce{YC7H13-Y$2$}, respectively, by the sensitivity analysis algorithm caused the most significant drops in post-ignition temperature.
In addition, the removal of reaction
\begin{equation*}
\cee{ACC6H10 + OH <-> ACC6H9-D + H2O} \tag{R3}
\end{equation*}
also contributed to the error at the lowest equivalence ratio.
Interestingly, removing only the above two species from the DRGEP-stage mechanism does not cause any change in post-ignition temperature, suggesting that the error accumulates during sensitivity analysis.
Nevertheless, the largest deviation in post-ignition temperature among all the conditions shown in \ref{F:hcci-temp-10atm} through \ref{F:hcci-temp-60atm} was \SI{3.5}{\percent} at the most.
Furthermore, even with these errors, both the skeletal and reduced mechanisms captured the two-stage ignition seen at lower temperatures for all three pressure\slash equivalence ratio combinations.

%% HCCI mechanism mixture variation
\begin{figure}[tbp]
	\centering
	\includegraphics[width=0.8\linewidth]{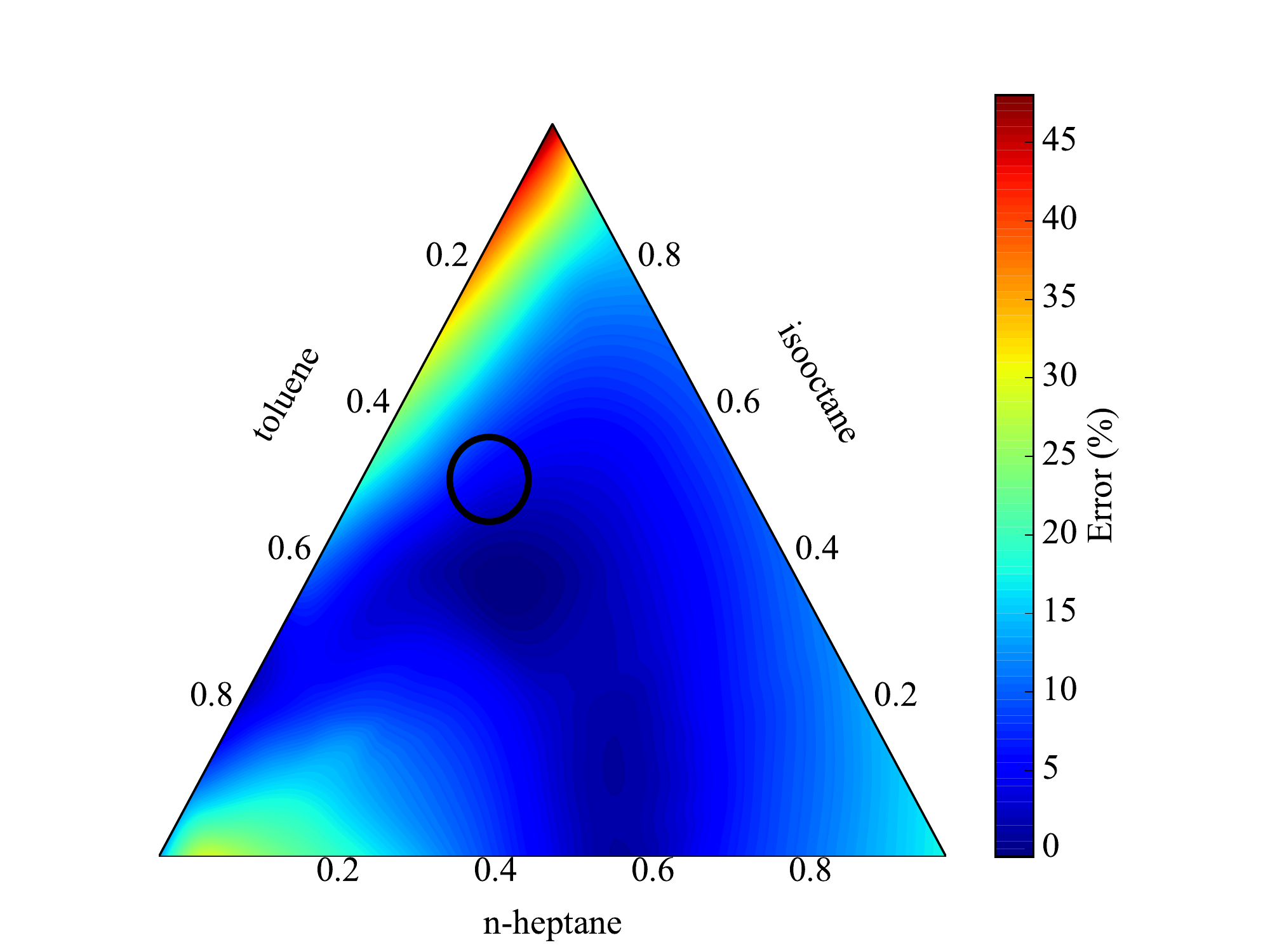}
	\caption{Percent error in predicted ignition delay for RON-like initial conditions (\SI{800}{\kelvin}, $\phi$ = 1.0, \SI{23}{\bar}) using the skeletal mechanism (213 species) targeted at HCCI-like conditions with varying mixture composition. The circle indicates the approximate location of the LLNL surrogate composition, with no 2-pentene content.}
	\label{F:hcci-mixture-variation}
\end{figure}

Next, we investigated the ability of the skeletal mechanism to predict autoignition for mixtures other than the LLNL surrogate, recognizing that acceptable performance is only guaranteed for the targeted conditions and mixture composition.
\ref{F:hcci-mixture-variation} shows the error in autoignition for RON-like initial conditions (\SI{800}{\kelvin}, $\phi$ = 1.0, \SI{23}{\bar}) with varying amounts of \emph{n}-heptane, isooctane, and toluene, a similar comparison to that in our previous study~\cite{Niemeyer:2014}.
In this case, since the amount of 2-pentene in the LLNL gasoline surrogate is small, we held the composition of 2-pentene at zero, in order to study the performance of the skeletal mechanism for TRF mixtures.
The skeletal mechanism performed reasonably well for a wide range of mixtures, with an average error of \SI{13.3}{\percent}; the error increased as the mixture approached the neat components, with a maximum of \SI{48.0}{\percent} at neat isooctane.
However, even with the acceptable performance for the particular conditions considered here, caution should be taken when applying the skeletal mechanism outside the target range of conditions~\cite{Niemeyer:2014}.

\subsubsection{SI\slash CI conditions}

\begin{figure}[tbp]
	\centering
	\includegraphics[width=0.7\linewidth]{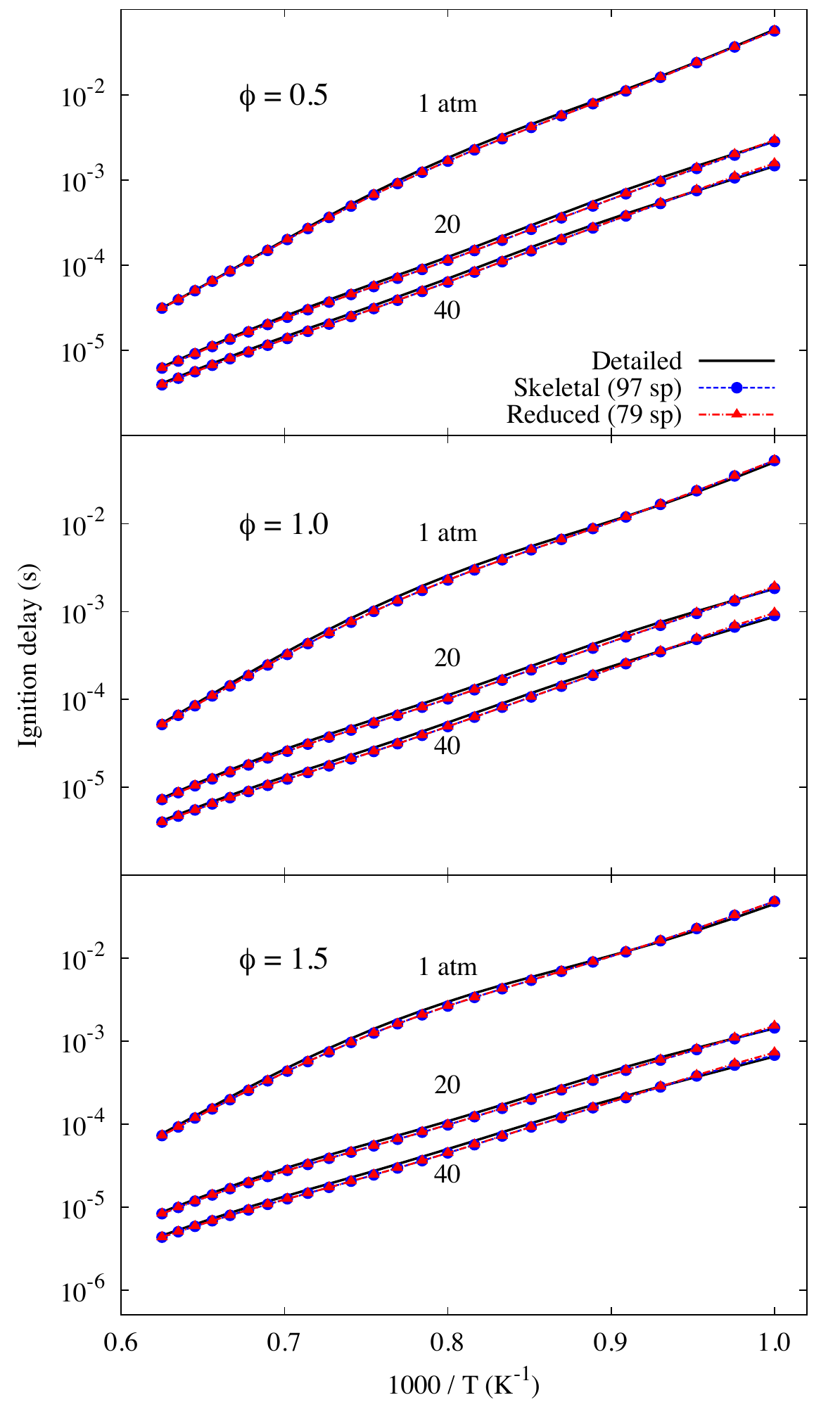}
	\caption{Autoignition validation of the skeletal (97 species) and reduced (79 species) mechanisms for the LLNL gasoline surrogate targeted at SI\slash CI-engine conditions.
	Ignition delay times were calculated over \SIrange{1000}{1600}{\kelvin}; \SIlist{1;20;40}{\atm}; and equivalence ratios of \numrange{0.5}{1.5} in air.}
	\label{F:hightemp-ign-delay}
\end{figure}

Next, we performed validation of the skeletal and reduced mechanisms for SI\slash CI conditions using constant-volume autoignition simulations over a range of initial conditions for temperature, pressure, and equivalence ratio.
\ref{F:hightemp-ign-delay} shows the comparison of the ignition delay times calculated by the detailed mechanism, skeletal mechanism resulting from the isomer lumping stage, and final reduced mechanism.
Both the skeletal and reduced mechanisms performed well over the full range of conditions considered, identically reproducing the ignition delays calculated by the detailed mechanism.

%% high-temp autoignition skeletal and reduced mechanisms

%% high-temp PSR skeletal and reduced mechanisms
\begin{figure}[tbp]
	\centering
	\includegraphics[width=0.9\linewidth]{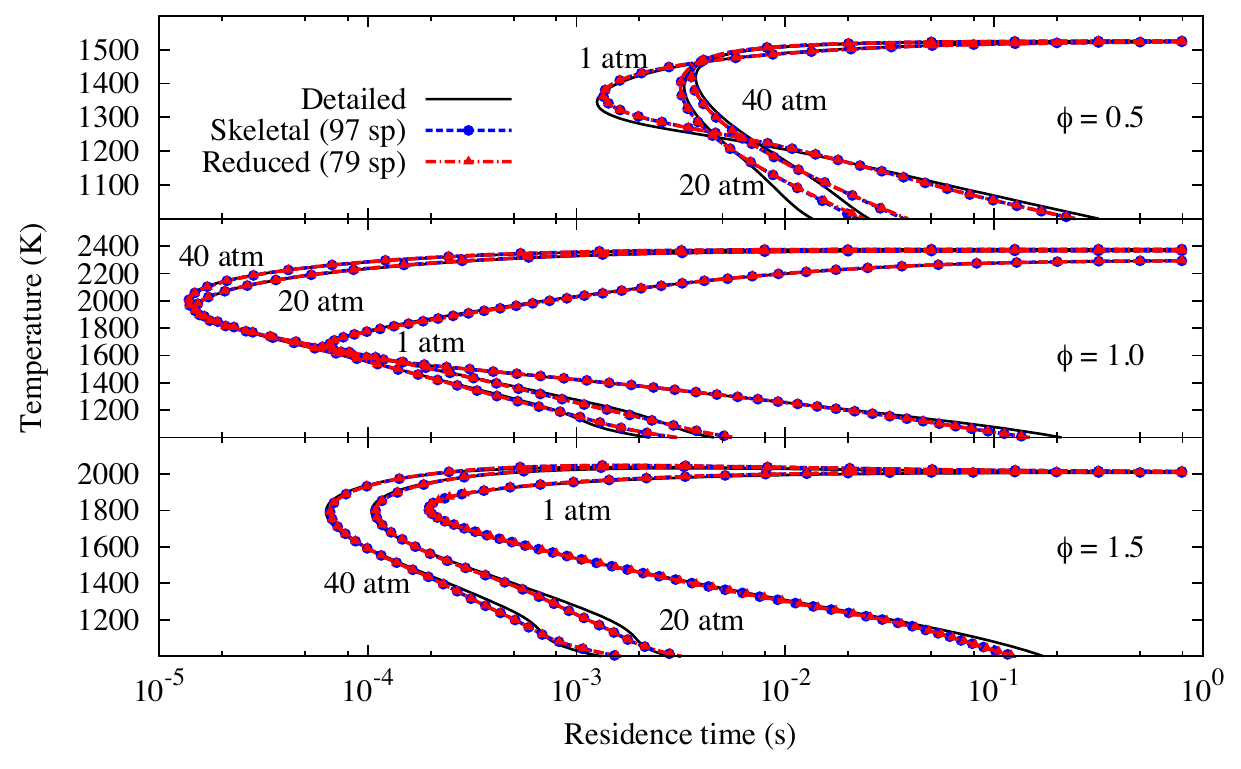}
	\caption{Comparison of PSR temperature response curves for the detailed, skeletal (97 species), and reduced (79 species) gasoline surrogate mechanisms generated using SI\slash CI conditions for pressures of \SIlist{1;20;40}{\atm}; equivalence ratios of \numrange{0.5}{1.5}; and an inlet temperature of \SI{300}{\kelvin}.}
	\label{F:hightemp-psr}
\end{figure}

% high-temp premix for greedy skeletal and reduced mechanisms
\begin{figure}[tbp]
	\centering
	\includegraphics[width=0.9\linewidth]{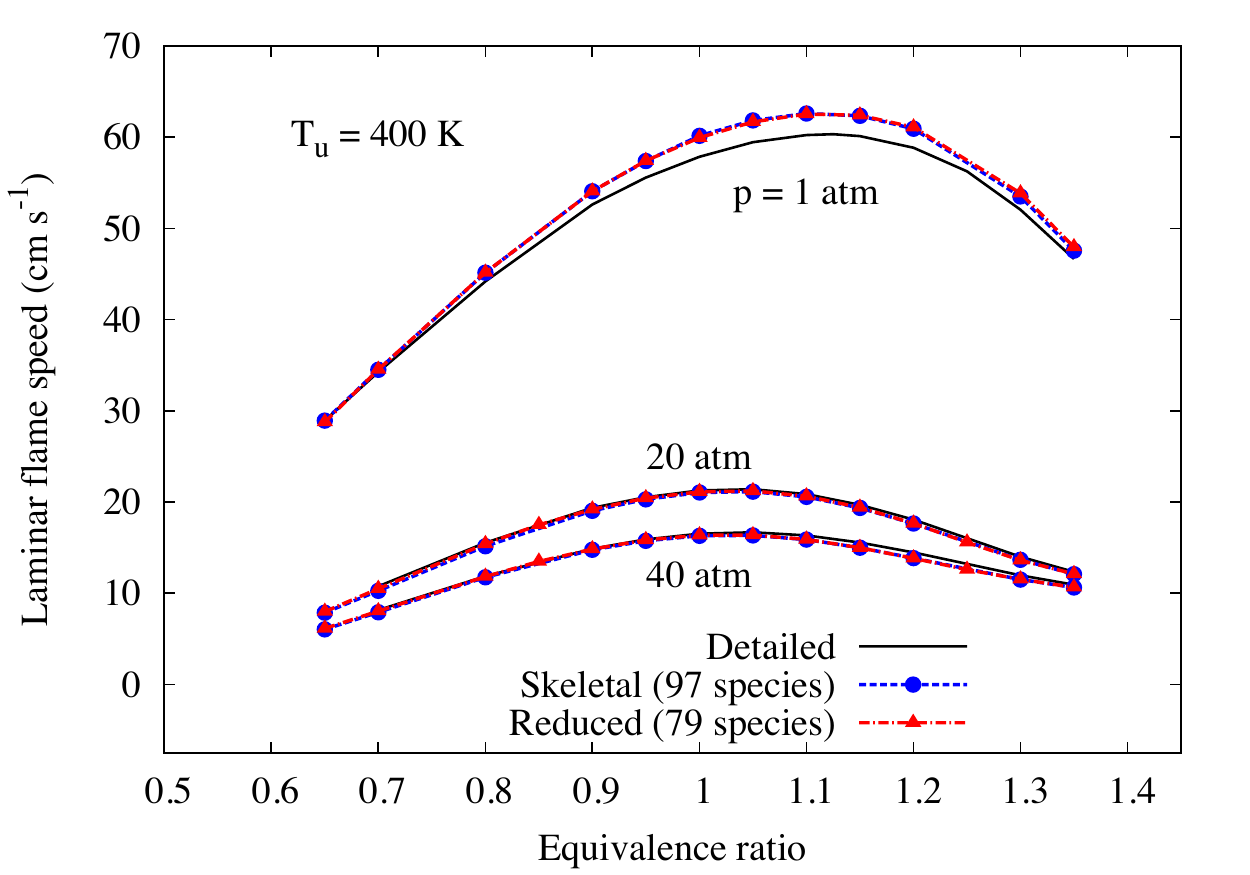}
	\caption{Comparison of laminar flame speeds calculated by the detailed, skeletal (97 species), and reduced (79 species) gasoline surrogate mechanisms generated using SI\slash CI conditions for a range of equivalence ratios at \SIlist{1;20;40}{\atm} and an unburned gas temperature $T_u =$ \SI{400}{\kelvin}.}
	\label{F:hightemp-flame-speed}
\end{figure}

In addition, we performed extended validation of the skeletal and reduced mechanisms using PSR and laminar flame speed simulations.
\ref{F:hightemp-psr} shows the PSR temperature response curves for pressures of \SIlist{1;20;40}{\atm}, equivalence ratios ranging over \numrange{0.5}{1.5}, and an inlet temperature of \SI{300}{\kelvin}.
Both the skeletal and reduced mechanisms reproduced the curves of the detailed mechanism for all conditions, and closely matched the extinction turning points.
In addition, \ref{F:hightemp-flame-speed} demonstrates the predictive ability of both the skeletal and reduced mechanisms in calculations of the laminar flame speeds, for pressures of \SIlist{1;20;40}{\atm}, equivalence ratios ranging over \numrange{0.65}{1.35}, and an unburned gas temperature of \SI{400}{\kelvin}.
Although some discrepancy is apparent near stoichiometric conditions for \SI{1}{\atm}, the maximum error is only \SI{4.9}{\percent} for $\phi = 1.0$; at higher pressures, the calculations of the skeletal and reduced mechanisms closely match those of the detailed mechanism.

%% high-temp mechanism mixture variation
\begin{figure}[tbp]
	\centering
	\includegraphics[width=0.8\linewidth]{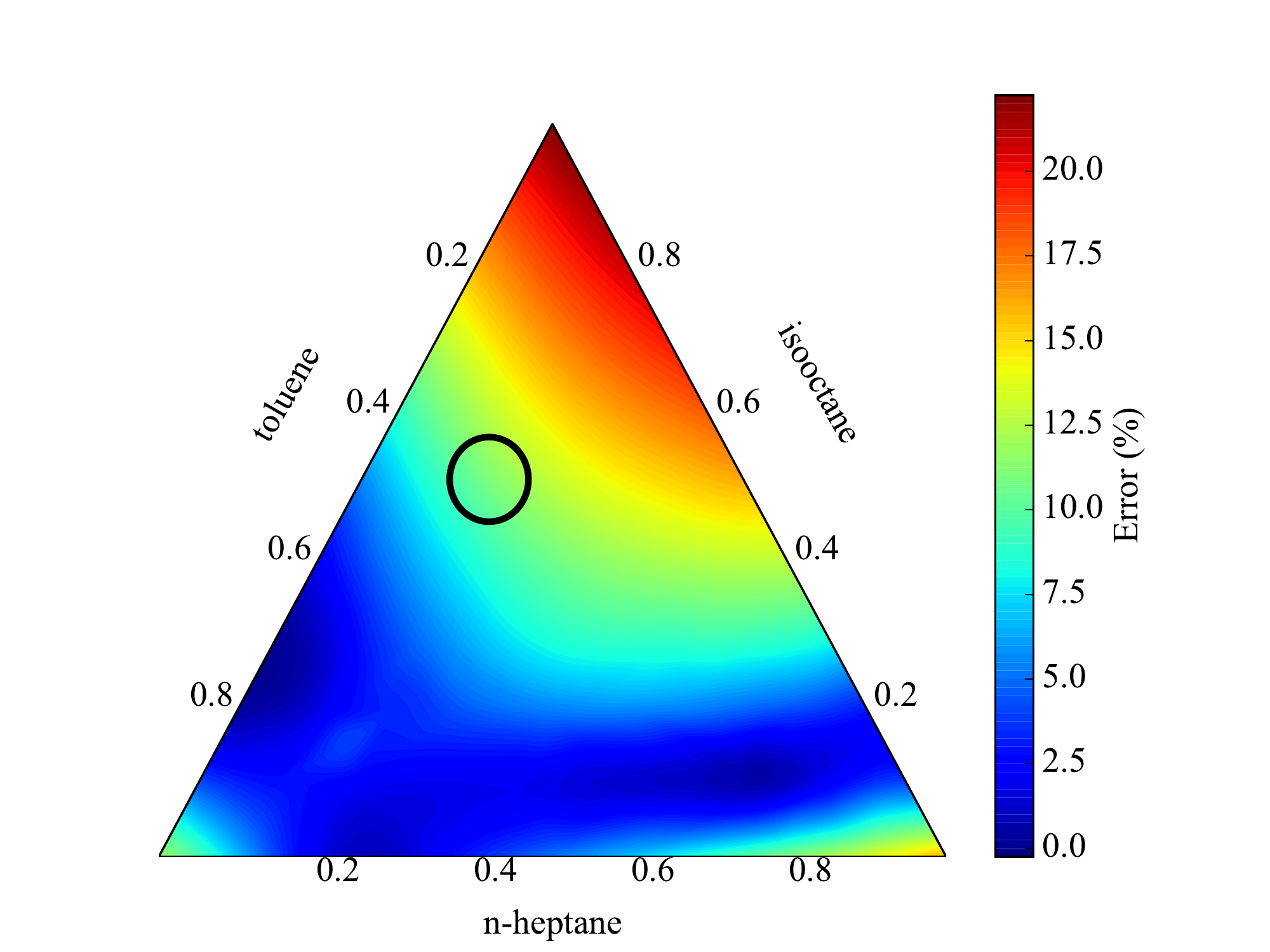}
	\caption{Percent error in predicted ignition delay for initial conditions of \SI{1200}{\kelvin}, $\phi$ = 1.0, \SI{10}{\atm} using the skeletal mechanism (97 species) targeted at SI\slash CI-like conditions with varying mixture composition. The circle indicates the approximate location of the LLNL surrogate composition, with no 2-pentene content.}
	\label{F:high-temp-mixture-variation}
\end{figure}

Finally, as with the HCCI-like mechanism, we also investigated the ability of the high-temperature skeletal mechanism to predict autoignition for varying mixture composition.
\ref{F:high-temp-mixture-variation} shows the error in autoignition at initial conditions of \SI{1200}{\kelvin}, $\phi$ = 1.0, and \SI{10}{\atm} with varying amounts of \emph{n}-heptane, isooctane, and toluene.
As before, we held the composition of 2-pentene at zero in order to study the performance of the skeletal mechanism for TRF mixtures.
The skeletal mechanism performed well for the entire mixture range, with an average error of \SI{9.20}{\percent} and maximum error of \SI{22.1}{\percent}.
As with the HCCI-like mechanism, here the error also increased as the mixture approached the neat components.
Again, we recommend caution when applying the skeletal mechanism outside the target range of conditions, as the performance may not be acceptable under other conditions~\cite{Niemeyer:2014}.

%%%%%%%%%%%%%%%%%%%%%%%%%%%%%%%%%%%%%%%%%%%%%%%%%%%%%%%%%%%%%%%%%%%%%%%%%%%%%%%%%%%%%%%%%%%
\section{Conclusions}

Skeletal and reduced mechanisms for the LLNL RD387 gasoline surrogate were generated using a combined strategy of skeletal reduction via the DRGEPSA method followed by further unimportant reaction elimination, isomer lumping, and finally time-scale reduction based on the QSS assumption using CSP analysis.
Starting with the original detailed mechanism with 1388 species and setting an error limit of \SI{10}{\percent}, skeletal and reduced mechanisms were generated with 213 and 148 species, respectively, for a lean-to-stoichiometric, low-temperature HCCI-like range of conditions, and 97 and 79 species, respectively, for a lean-to-rich, high-temperature SI\slash CI-like range of conditions.
Validation of the resulting skeletal and reduced mechanisms showed good performance in predicting homogenous autoignition delay over the appropriate ranges of conditions.
For the HCCI-like range of conditions, the associated skeletal and reduced mechanisms also performed well in more rigorous validation studies using temperature profiles in constant-volume autoignition simulations.
Similarly, the SI\slash CI skeletal and reduced mechanisms closely matched the PSR temperature response curves and extinction turning points as well as the laminar flame speeds of the starting detailed mechanism.

For the high-temperature SI\slash CI range of conditions, the final skeletal and reduced mechanisms are much more compact for consideration in multidimensional engine simulations while still capable of predicting relevant combustion properties nearly as well as the much larger detailed mechanism.
However, the skeletal and reduced mechanisms produced here may still be too large to use in high-fidelity reactive flow simulations, even considering the significant extent of reduction compared to the original detailed mechanism.
Therefore, while incorporating reduced chemistry in multidimensional reactive-flow simulations is the focus of our future work, dynamic mechanism reduction techniques may be required and are also under investigation~\cite{Curtis:2014aa}.

In addition to the skeletal and reduced mechanisms produced through the course of this study, improvements and observations were made regarding the reduction procedure itself.
A new greedy sensitivity analysis algorithm for DRGEPSA was demonstrated, and shown to be capable of eliminating a larger number of species for the same error limit as the previous algorithm.
Furthermore, we found that this algorithm, combined with the addition of PSR data to the reduction process and the use of a high value for the sensitivity analysis cutoff threshold, produced a notably smaller skeletal mechanism that also performed well in autoignition, PSR, and laminar flame speed calculations.
While many of the parameters involved in our reduction procedure are determined automatically based on the set error limit, it does involve some other user-defined values especially during the time-scale reduction stage.
Based on the mechanism reductions performed in this work, we can make several suggestions for setting these values:
\begin{itemize}
\item The upper threshold for sensitivity analysis can be set to a high value, such as 0.5, when using the greedy algorithm combined with autoignition and PSR data.
\item The CSP characteristic time scale safety factor should be set to a value (e.g., 100) large enough to significantly separate the time scales of the fast modes from that of the system.
\item The CSP cutoff threshold should be set at a lower value, such as \num{1e-4}, than indicated previously~\cite{Lu:2008dp,Lu:2008bi}.
\item The QSS graph cutoff threshold should be set to a small but nonzero value such as 0.01 to sufficiently trim the graph to prevent extremely long and complex expressions in the analytic QSS species concentration solution.
\end{itemize}
However, additional investigation is warranted for the development of an automated, robust selection approach for the optimal CSP cutoff threshold based on, e.g., the maximum contributions of the nonlinear QSS terms; the ad hoc, manual selection method used previously may result in a poorly performing reduced mechanism under certain conditions.

%%%%%%%%%%%%%%%%%%%%%%%%%%%%%%%%%%%%%%%%%%%%%%%%%%%%%%%%%%%%%%%%%%%%%%%%%%%%%%%%%%%%%%%%%%%
\acknowledgement

This work was partially supported by the National Science Foundation under Grant No.\ CBET-1402231 and the National Science Foundation Graduate Research Fellowship under Grant No.\ 0951783.

%%%%%%%%%%%%%%%%%%%%%%%%%%%%%%%%%%%%%%%%%%%%%%%%%%%%%%%%%%%%%%%%%%%%%%%%%%%%%%%%%%%%%%%%%%%
\begin{suppinfo}
The skeletal and reduced mechanisms for the gasoline surrogate generated for both HCCI and SI\slash CI engine conditions will be made available as supplementary material upon acceptance for publication.
\end{suppinfo}

\providecommand{\latin}[1]{#1}
\providecommand*\mcitethebibliography{\thebibliography}
\csname @ifundefined\endcsname{endmcitethebibliography}
  {\let\endmcitethebibliography\endthebibliography}{}


\begin{mcitethebibliography}{50}
\providecommand*\natexlab[1]{#1}
\providecommand*\mciteSetBstSublistMode[1]{}
\providecommand*\mciteSetBstMaxWidthForm[2]{}
\providecommand*\mciteBstWouldAddEndPuncttrue
  {\def\EndOfBibitem{\unskip.}}
\providecommand*\mciteBstWouldAddEndPunctfalse
  {\let\EndOfBibitem\relax}
\providecommand*\mciteSetBstMidEndSepPunct[3]{}
\providecommand*\mciteSetBstSublistLabelBeginEnd[3]{}
\providecommand*\EndOfBibitem{}
\mciteSetBstSublistMode{f}
\mciteSetBstMaxWidthForm{subitem}{(\alph{mcitesubitemcount})}
\mciteSetBstSublistLabelBeginEnd
  {\mcitemaxwidthsubitemform\space}
  {\relax}
  {\relax}

\bibitem[Edgar(1927)]{Edgar:1927fq}
Edgar,~G. Measurement of knock characteristics of gasoline in terms of a
  standard fuel. \emph{Ind. Eng. Chem.} \textbf{1927}, \emph{19},
  145--146\relax
\mciteBstWouldAddEndPuncttrue
\mciteSetBstMidEndSepPunct{\mcitedefaultmidpunct}
{\mcitedefaultendpunct}{\mcitedefaultseppunct}\relax
\EndOfBibitem
\bibitem[Sturgis \latin{et~al.}(1954)Sturgis, Cantwell, Morris, and
  Schultz]{Sturgis:1954uq}
Sturgis,~B.~M.; Cantwell,~E.~N.; Morris,~W.~E.; Schultz,~D.~L. The preignition
  resistance of fuels. \emph{Proc. API} \textbf{1954}, \emph{34},
  256--269\relax
\mciteBstWouldAddEndPuncttrue
\mciteSetBstMidEndSepPunct{\mcitedefaultmidpunct}
{\mcitedefaultendpunct}{\mcitedefaultseppunct}\relax
\EndOfBibitem
\bibitem[Pahnke \latin{et~al.}(1954)Pahnke, Cohen, and Sturgis]{Pahnke:1954er}
Pahnke,~A.~J.; Cohen,~P.~M.; Sturgis,~B.~M. Preflame oxidation of hydrocarbons
  in a motored engine. \emph{Ind. Eng. Chem.} \textbf{1954}, \emph{46},
  1024--1029\relax
\mciteBstWouldAddEndPuncttrue
\mciteSetBstMidEndSepPunct{\mcitedefaultmidpunct}
{\mcitedefaultendpunct}{\mcitedefaultseppunct}\relax
\EndOfBibitem
\bibitem[Chaos \latin{et~al.}(2007)Chaos, Zhao, Kazakov, Gokulakrishnan,
  Angioletti, and Dryer]{Chaos:2007}
Chaos,~M.; Zhao,~Z.; Kazakov,~A.; Gokulakrishnan,~P.; Angioletti,~M.; Dryer,~F.
  A {PRF}+toluene surrogate fuel model for simulating gasoline kinetics. 5th US
  Combustion Meeting. 2007\relax
\mciteBstWouldAddEndPuncttrue
\mciteSetBstMidEndSepPunct{\mcitedefaultmidpunct}
{\mcitedefaultendpunct}{\mcitedefaultseppunct}\relax
\EndOfBibitem
\bibitem[Gauthier \latin{et~al.}(2004)Gauthier, Davidson, and
  Hanson]{Gauthier:2004}
Gauthier,~B.~M.; Davidson,~D.~F.; Hanson,~R.~K. Shock tube determination of
  ignition delay times in full-blend and surrogate fuel mixtures.
  \emph{Combust. Flame} \textbf{2004}, \emph{139}, 300--311\relax
\mciteBstWouldAddEndPuncttrue
\mciteSetBstMidEndSepPunct{\mcitedefaultmidpunct}
{\mcitedefaultendpunct}{\mcitedefaultseppunct}\relax
\EndOfBibitem
\bibitem[Mehl \latin{et~al.}(2011)Mehl, Pitz, Westbrook, and
  Curran]{Mehl:2011cn}
Mehl,~M.; Pitz,~W.~J.; Westbrook,~C.~K.; Curran,~H.~J. Kinetic modeling of
  gasoline surrogate components and mixtures under engine conditions.
  \emph{Proc. Combust. Inst.} \textbf{2011}, \emph{33}, 193--200\relax
\mciteBstWouldAddEndPuncttrue
\mciteSetBstMidEndSepPunct{\mcitedefaultmidpunct}
{\mcitedefaultendpunct}{\mcitedefaultseppunct}\relax
\EndOfBibitem
\bibitem[Mehl \latin{et~al.}(2011)Mehl, Chen, Pitz, Sarathy, and
  Westbrook]{Mehl:2011jn}
Mehl,~M.; Chen,~J.-Y.; Pitz,~W.~J.; Sarathy,~S.~M.; Westbrook,~C.~K. An
  approach for formulating surrogates for gasoline with application toward a
  reduced surrogate mechanism for {CFD} engine modeling. \emph{Energy Fuels}
  \textbf{2011}, \emph{25}, 5215--5223\relax
\mciteBstWouldAddEndPuncttrue
\mciteSetBstMidEndSepPunct{\mcitedefaultmidpunct}
{\mcitedefaultendpunct}{\mcitedefaultseppunct}\relax
\EndOfBibitem
\bibitem[Kukkadapu \latin{et~al.}(2012)Kukkadapu, Kumar, Sung, Mehl, and
  Pitz]{Kukkadapu:2012dx}
Kukkadapu,~G.; Kumar,~K.; Sung,~C.~J.; Mehl,~M.; Pitz,~W.~J. Experimental and
  surrogate modeling study of gasoline ignition in a rapid compression machine.
  \emph{Combust. Flame} \textbf{2012}, \emph{159}, 3066--3078\relax
\mciteBstWouldAddEndPuncttrue
\mciteSetBstMidEndSepPunct{\mcitedefaultmidpunct}
{\mcitedefaultendpunct}{\mcitedefaultseppunct}\relax
\EndOfBibitem
\bibitem[Kukkadapu \latin{et~al.}(2013)Kukkadapu, Kumar, Sung, Mehl, and
  Pitz]{Kukkadapu:2013ko}
Kukkadapu,~G.; Kumar,~K.; Sung,~C.~J.; Mehl,~M.; Pitz,~W.~J. Autoignition of
  gasoline and its surrogates in a rapid compression machine. \emph{Proc.
  Combust. Inst.} \textbf{2013}, \emph{34}, 345--352\relax
\mciteBstWouldAddEndPuncttrue
\mciteSetBstMidEndSepPunct{\mcitedefaultmidpunct}
{\mcitedefaultendpunct}{\mcitedefaultseppunct}\relax
\EndOfBibitem
\bibitem[Sarathy \latin{et~al.}(2015)Sarathy, Kukkadapu, Mehl, Wang, Javed,
  Park, Oehlschlaeger, Farooq, Pitz, and Sung]{Sarathy:2014aa}
Sarathy,~S.~M.; Kukkadapu,~G.; Mehl,~M.; Wang,~W.; Javed,~T.; Park,~W.;
  Oehlschlaeger,~M.~A.; Farooq,~A.; Pitz,~W.~J.; Sung,~C.~J. Ignition of
  alkane-rich {FACE} gasoline fuels and their surrogate mixtures. \emph{Proc.
  Combust. Inst.} \textbf{2015}, \emph{35}, 249--257\relax
\mciteBstWouldAddEndPuncttrue
\mciteSetBstMidEndSepPunct{\mcitedefaultmidpunct}
{\mcitedefaultendpunct}{\mcitedefaultseppunct}\relax
\EndOfBibitem
\bibitem[Lu and Law(2009)Lu, and Law]{Lu:2009gh}
Lu,~T.; Law,~C.~K. Toward accommodating realistic fuel chemistry in large-scale
  computations. \emph{Prog. Energy Comb. Sci.} \textbf{2009}, \emph{35},
  192--215\relax
\mciteBstWouldAddEndPuncttrue
\mciteSetBstMidEndSepPunct{\mcitedefaultmidpunct}
{\mcitedefaultendpunct}{\mcitedefaultseppunct}\relax
\EndOfBibitem
\bibitem[Lu and Law(2005)Lu, and Law]{Lu:2005ce}
Lu,~T.; Law,~C.~K. A directed relation graph method for mechanism reduction.
  \emph{Proc. Combust. Inst.} \textbf{2005}, \emph{30}, 1333--1341\relax
\mciteBstWouldAddEndPuncttrue
\mciteSetBstMidEndSepPunct{\mcitedefaultmidpunct}
{\mcitedefaultendpunct}{\mcitedefaultseppunct}\relax
\EndOfBibitem
\bibitem[Lu and Law(2006)Lu, and Law]{Lu:2006bb}
Lu,~T.; Law,~C.~K. Linear time reduction of large kinetic mechanisms with
  directed relation graph: \emph{n}-heptane and iso-octane. \emph{Combust.
  Flame} \textbf{2006}, \emph{144}, 24--36\relax
\mciteBstWouldAddEndPuncttrue
\mciteSetBstMidEndSepPunct{\mcitedefaultmidpunct}
{\mcitedefaultendpunct}{\mcitedefaultseppunct}\relax
\EndOfBibitem
\bibitem[Lu and Law(2006)Lu, and Law]{Lu:2006gi}
Lu,~T.; Law,~C.~K. On the applicability of directed relation graphs to the
  reduction of reaction mechanisms. \emph{Combust. Flame} \textbf{2006},
  \emph{146}, 472--483\relax
\mciteBstWouldAddEndPuncttrue
\mciteSetBstMidEndSepPunct{\mcitedefaultmidpunct}
{\mcitedefaultendpunct}{\mcitedefaultseppunct}\relax
\EndOfBibitem
\bibitem[Bendtsen \latin{et~al.}(2001)Bendtsen, Glarborg, and
  Dam-Johansen]{Bendtsen:2001vh}
Bendtsen,~A.; Glarborg,~P.; Dam-Johansen,~K. Visualization methods in analysis
  of detailed chemical kinetics modelling. \emph{Comput. Chem.} \textbf{2001},
  \emph{25}, 161--170\relax
\mciteBstWouldAddEndPuncttrue
\mciteSetBstMidEndSepPunct{\mcitedefaultmidpunct}
{\mcitedefaultendpunct}{\mcitedefaultseppunct}\relax
\EndOfBibitem
\bibitem[Sankaran \latin{et~al.}(2007)Sankaran, Hawkes, Chen, Lu, and
  Law]{Sankaran:2007fs}
Sankaran,~R.; Hawkes,~E.~R.; Chen,~J.~H.; Lu,~T.; Law,~C.~K. Structure of a
  spatially developing turbulent lean methane-air {Bunsen} flame. \emph{Proc.
  Combust. Inst.} \textbf{2007}, \emph{31}, 1291--1298\relax
\mciteBstWouldAddEndPuncttrue
\mciteSetBstMidEndSepPunct{\mcitedefaultmidpunct}
{\mcitedefaultendpunct}{\mcitedefaultseppunct}\relax
\EndOfBibitem
\bibitem[Zheng \latin{et~al.}(2007)Zheng, Lu, and Law]{Zheng:2007gd}
Zheng,~X.~L.; Lu,~T.; Law,~C.~K. Experimental counterflow ignition temperatures
  and reaction mechanisms of 1,3-butadiene. \emph{Proc. Combust. Inst.}
  \textbf{2007}, \emph{31}, 367--375\relax
\mciteBstWouldAddEndPuncttrue
\mciteSetBstMidEndSepPunct{\mcitedefaultmidpunct}
{\mcitedefaultendpunct}{\mcitedefaultseppunct}\relax
\EndOfBibitem
\bibitem[Lu and Law(2008)Lu, and Law]{Lu:2008bi}
Lu,~T.; Law,~C.~K. Strategies for mechanism reduction for large hydrocarbons:
  n-heptane. \emph{Combust. Flame} \textbf{2008}, \emph{154}, 153--163\relax
\mciteBstWouldAddEndPuncttrue
\mciteSetBstMidEndSepPunct{\mcitedefaultmidpunct}
{\mcitedefaultendpunct}{\mcitedefaultseppunct}\relax
\EndOfBibitem
\bibitem[Pepiot-Desjardins and Pitsch(2008)Pepiot-Desjardins, and
  Pitsch]{Pepiot-Desjardins:2008}
Pepiot-Desjardins,~P.; Pitsch,~H. An efficient error-propagation-based
  reduction method for large chemical kinetic mechanisms. \emph{Combust. Flame}
  \textbf{2008}, \emph{154}, 67--81\relax
\mciteBstWouldAddEndPuncttrue
\mciteSetBstMidEndSepPunct{\mcitedefaultmidpunct}
{\mcitedefaultendpunct}{\mcitedefaultseppunct}\relax
\EndOfBibitem
\bibitem[Niemeyer and Sung(2011)Niemeyer, and Sung]{Niemeyer:2011fe}
Niemeyer,~K.~E.; Sung,~C.~J. On the importance of graph search algorithms for
  {DRGEP}-based mechanism reduction methods. \emph{Combust. Flame}
  \textbf{2011}, \emph{158}, 1439--1443\relax
\mciteBstWouldAddEndPuncttrue
\mciteSetBstMidEndSepPunct{\mcitedefaultmidpunct}
{\mcitedefaultendpunct}{\mcitedefaultseppunct}\relax
\EndOfBibitem
\bibitem[Niemeyer \latin{et~al.}(2010)Niemeyer, Sung, and
  Raju]{Niemeyer:2010bt}
Niemeyer,~K.~E.; Sung,~C.~J.; Raju,~M.~P. Skeletal mechanism generation for
  surrogate fuels using directed relation graph with error propagation and
  sensitivity analysis. \emph{Combust. Flame} \textbf{2010}, \emph{157},
  1760--1770\relax
\mciteBstWouldAddEndPuncttrue
\mciteSetBstMidEndSepPunct{\mcitedefaultmidpunct}
{\mcitedefaultendpunct}{\mcitedefaultseppunct}\relax
\EndOfBibitem
\bibitem[Niemeyer and Sung(2014)Niemeyer, and Sung]{Niemeyer:2014}
Niemeyer,~K.~E.; Sung,~C.~J. Mechanism reduction for multicomponent surrogates:
  A case study using toluene reference fuels. \emph{Combust. Flame}
  \textbf{2014}, \emph{161}, 2752--2764\relax
\mciteBstWouldAddEndPuncttrue
\mciteSetBstMidEndSepPunct{\mcitedefaultmidpunct}
{\mcitedefaultendpunct}{\mcitedefaultseppunct}\relax
\EndOfBibitem
\bibitem[Sun \latin{et~al.}(2010)Sun, Chen, Gou, and Ju]{Sun:2010jf}
Sun,~W.; Chen,~Z.; Gou,~X.; Ju,~Y. A path flux analysis method for the
  reduction of detailed chemical kinetic mechanisms. \emph{Combust. Flame}
  \textbf{2010}, \emph{157}, 1298--1307\relax
\mciteBstWouldAddEndPuncttrue
\mciteSetBstMidEndSepPunct{\mcitedefaultmidpunct}
{\mcitedefaultendpunct}{\mcitedefaultseppunct}\relax
\EndOfBibitem
\bibitem[Bodenstein(1913)]{Bodenstein:1913tc}
Bodenstein,~M. Eine theorie der photomechnischen reaktionsgeschwindigkeit.
  \emph{Z. Phys. Chem.} \textbf{1913}, \emph{85}, 329--397\relax
\mciteBstWouldAddEndPuncttrue
\mciteSetBstMidEndSepPunct{\mcitedefaultmidpunct}
{\mcitedefaultendpunct}{\mcitedefaultseppunct}\relax
\EndOfBibitem
\bibitem[Chapman and Underhill(1913)Chapman, and Underhill]{Chapman:1913dx}
Chapman,~D.; Underhill,~L. {LV}.---The interaction of chlorine and hydrogen.
  The influence of mass. \emph{J. Chem. Soc., Trans.} \textbf{1913},
  \emph{103}, 496--508\relax
\mciteBstWouldAddEndPuncttrue
\mciteSetBstMidEndSepPunct{\mcitedefaultmidpunct}
{\mcitedefaultendpunct}{\mcitedefaultseppunct}\relax
\EndOfBibitem
\bibitem[Benson(1952)]{Benson:1952ju}
Benson,~S.~W. The induction period in chain reactions. \emph{J. Chem. Phys.}
  \textbf{1952}, \emph{20}, 1605--1612\relax
\mciteBstWouldAddEndPuncttrue
\mciteSetBstMidEndSepPunct{\mcitedefaultmidpunct}
{\mcitedefaultendpunct}{\mcitedefaultseppunct}\relax
\EndOfBibitem
\bibitem[Ramshaw(1980)]{Ramshaw:1980kn}
Ramshaw,~J.~D. Partial equilibrium in fluid dynamics. \emph{Phys. Fluids}
  \textbf{1980}, \emph{23}, 675--680\relax
\mciteBstWouldAddEndPuncttrue
\mciteSetBstMidEndSepPunct{\mcitedefaultmidpunct}
{\mcitedefaultendpunct}{\mcitedefaultseppunct}\relax
\EndOfBibitem
\bibitem[Lam and Goussis(1988)Lam, and Goussis]{Lam:1988wc}
Lam,~S.~H.; Goussis,~D.~A. Understanding complex chemical kinetics with
  computational singular perturbation. \emph{Proc. Combust. Inst.}
  \textbf{1988}, \emph{22}, 931--941\relax
\mciteBstWouldAddEndPuncttrue
\mciteSetBstMidEndSepPunct{\mcitedefaultmidpunct}
{\mcitedefaultendpunct}{\mcitedefaultseppunct}\relax
\EndOfBibitem
\bibitem[Lam(1993)]{Lam:1993ub}
Lam,~S.~H. Using {CSP} to understand complex chemical kinetics. \emph{Combust.
  Sci. Technol.} \textbf{1993}, \emph{89}, 375--404\relax
\mciteBstWouldAddEndPuncttrue
\mciteSetBstMidEndSepPunct{\mcitedefaultmidpunct}
{\mcitedefaultendpunct}{\mcitedefaultseppunct}\relax
\EndOfBibitem
\bibitem[Lam and Goussis(1994)Lam, and Goussis]{Lam:1994ws}
Lam,~S.~H.; Goussis,~D.~A. The {CSP} method for simplying kinetics. \emph{Int.
  J. Chem. Kinet.} \textbf{1994}, \emph{26}, 461--486\relax
\mciteBstWouldAddEndPuncttrue
\mciteSetBstMidEndSepPunct{\mcitedefaultmidpunct}
{\mcitedefaultendpunct}{\mcitedefaultseppunct}\relax
\EndOfBibitem
\bibitem[Maas and Pope(1992)Maas, and Pope]{Maas:1992ws}
Maas,~U.; Pope,~S.~B. Simplifying chemical kinetics: intrinsic low-dimensional
  manifolds in composition space. \emph{Combust. Flame} \textbf{1992},
  \emph{88}, 239--264\relax
\mciteBstWouldAddEndPuncttrue
\mciteSetBstMidEndSepPunct{\mcitedefaultmidpunct}
{\mcitedefaultendpunct}{\mcitedefaultseppunct}\relax
\EndOfBibitem
\bibitem[Niemeyer(2010)]{Niemeyer:2010}
Niemeyer,~K.~E. Skeletal mechanism generation for surrogate fuels. {MS} Thesis,
  Case Western Reserve University, 2010\relax
\mciteBstWouldAddEndPuncttrue
\mciteSetBstMidEndSepPunct{\mcitedefaultmidpunct}
{\mcitedefaultendpunct}{\mcitedefaultseppunct}\relax
\EndOfBibitem
\bibitem[{OpenMP Architecture Review Board}(2008)]{OpenMP:2008}
{OpenMP Architecture Review Board}, {OpenMP} Application Program Interface
  Version 3.0. \url{http://www.openmp.org/mp-documents/spec30.pdf}, 2008\relax
\mciteBstWouldAddEndPuncttrue
\mciteSetBstMidEndSepPunct{\mcitedefaultmidpunct}
{\mcitedefaultendpunct}{\mcitedefaultseppunct}\relax
\EndOfBibitem
\bibitem[Cormen \latin{et~al.}(2009)Cormen, Leiserson, Rivest, and
  Stein]{Cormen:2009uw}
Cormen,~T.~H.; Leiserson,~C.~E.; Rivest,~R.~L.; Stein,~C. \emph{Introduction to
  Algorithms}, 3rd ed.; The MIT Press: Cambridge, Massachusetts, 2009\relax
\mciteBstWouldAddEndPuncttrue
\mciteSetBstMidEndSepPunct{\mcitedefaultmidpunct}
{\mcitedefaultendpunct}{\mcitedefaultseppunct}\relax
\EndOfBibitem
\bibitem[Ahmed \latin{et~al.}(2007)Ahmed, Mau{\ss}, Mor{\'e}ac, and
  Zeuch]{Ahmed:2007fa}
Ahmed,~S.~S.; Mau{\ss},~F.; Mor{\'e}ac,~G.; Zeuch,~T. A comprehensive and
  compact \emph{n}-heptane oxidation model derived using chemical lumping.
  \emph{Phys. Chem. Chem. Phys.} \textbf{2007}, \emph{9}, 1107--1126\relax
\mciteBstWouldAddEndPuncttrue
\mciteSetBstMidEndSepPunct{\mcitedefaultmidpunct}
{\mcitedefaultendpunct}{\mcitedefaultseppunct}\relax
\EndOfBibitem
\bibitem[Pepiot-Desjardins and Pitsch(2008)Pepiot-Desjardins, and
  Pitsch]{Pepiot:2008kq}
Pepiot-Desjardins,~P.; Pitsch,~H. An automatic chemical lumping method for the
  reduction of large chemical kinetic mechanisms. \emph{Combust. Theor. Model.}
  \textbf{2008}, \emph{12}, 1089--1108\relax
\mciteBstWouldAddEndPuncttrue
\mciteSetBstMidEndSepPunct{\mcitedefaultmidpunct}
{\mcitedefaultendpunct}{\mcitedefaultseppunct}\relax
\EndOfBibitem
\bibitem[Goussis and Lam(1992)Goussis, and Lam]{Goussis:1992ez}
Goussis,~D.~A.; Lam,~S.~H. A study of homogeneous methanol oxidation kinetics
  using {CSP}. \emph{Proc. Combust. Inst.} \textbf{1992}, \emph{24},
  113--120\relax
\mciteBstWouldAddEndPuncttrue
\mciteSetBstMidEndSepPunct{\mcitedefaultmidpunct}
{\mcitedefaultendpunct}{\mcitedefaultseppunct}\relax
\EndOfBibitem
\bibitem[Lu and Law(2008)Lu, and Law]{Lu:2008dp}
Lu,~T.; Law,~C.~K. A criterion based on computational singular perturbation for
  the identification of quasi steady state species: A reduced mechanism for
  methane oxidation with {NO} chemistry. \emph{Combust. Flame} \textbf{2008},
  \emph{154}, 761--774\relax
\mciteBstWouldAddEndPuncttrue
\mciteSetBstMidEndSepPunct{\mcitedefaultmidpunct}
{\mcitedefaultendpunct}{\mcitedefaultseppunct}\relax
\EndOfBibitem
\bibitem[Anderson \latin{et~al.}(1999)Anderson, Bai, Bischof, Blackford,
  Demmel, Dongarra, Du~Croz, Greenbaum, Hammarling, McKenney, and
  Sorensen]{Anderson:1999}
Anderson,~E.; Bai,~Z.; Bischof,~C.; Blackford,~S.; Demmel,~J.; Dongarra,~J.;
  Du~Croz,~J.; Greenbaum,~A.; Hammarling,~S.; McKenney,~A.; Sorensen,~D.
  \emph{{LAPACK} Users' Guide}, 3rd ed.; Society for Industrial and Applied
  Mathematics: Philadelphia, PA, 1999\relax
\mciteBstWouldAddEndPuncttrue
\mciteSetBstMidEndSepPunct{\mcitedefaultmidpunct}
{\mcitedefaultendpunct}{\mcitedefaultseppunct}\relax
\EndOfBibitem
\bibitem[Law \latin{et~al.}(2003)Law, Sung, Wang, and Lu]{Law:2003wt}
Law,~C.~K.; Sung,~C.~J.; Wang,~H.; Lu,~T. Development of comprehensive detailed
  and reduced reaction mechanisms for combustion modeling. \emph{AIAA J.}
  \textbf{2003}, \emph{41}, 1629--1646\relax
\mciteBstWouldAddEndPuncttrue
\mciteSetBstMidEndSepPunct{\mcitedefaultmidpunct}
{\mcitedefaultendpunct}{\mcitedefaultseppunct}\relax
\EndOfBibitem
\bibitem[Lu and Law(2006)Lu, and Law]{Lu:2006cn}
Lu,~T.; Law,~C.~K. Systematic approach to obtain analytic solutions of quasi
  steady state species in reduced mechanisms. \emph{J. Phys. Chem. A}
  \textbf{2006}, \emph{110}, 13202--13208\relax
\mciteBstWouldAddEndPuncttrue
\mciteSetBstMidEndSepPunct{\mcitedefaultmidpunct}
{\mcitedefaultendpunct}{\mcitedefaultseppunct}\relax
\EndOfBibitem
\bibitem[Burkardt(2006)]{Burkardt:2006}
Burkardt,~J. {GRAFPACK}.
  \url{http://people.sc.fsu.edu/~jburkardt/f_src/grafpack/grafpack.html },
  2006\relax
\mciteBstWouldAddEndPuncttrue
\mciteSetBstMidEndSepPunct{\mcitedefaultmidpunct}
{\mcitedefaultendpunct}{\mcitedefaultseppunct}\relax
\EndOfBibitem
\bibitem[Thulasiraman and Swamy(1992)Thulasiraman, and
  Swamy]{Thulasiraman:1992}
Thulasiraman,~K.; Swamy,~M. N.~S. \emph{Graphs: Theory and Algorithms}; John
  Wiley \& Sons, Inc.: New York, 1992\relax
\mciteBstWouldAddEndPuncttrue
\mciteSetBstMidEndSepPunct{\mcitedefaultmidpunct}
{\mcitedefaultendpunct}{\mcitedefaultseppunct}\relax
\EndOfBibitem
\bibitem[Kee \latin{et~al.}(1996)Kee, Rupley, Meeks, and Miller]{Kee:1996vd}
Kee,~R.~J.; Rupley,~F.~M.; Meeks,~E.; Miller,~J.~A. {CHEMKIN-III}: a {FORTRAN}
  chemical kinetics package for the analysis of gas-phase chemical and plasma
  kinetics. Sandia National Laboratories, SAND96-8216, 1996\relax
\mciteBstWouldAddEndPuncttrue
\mciteSetBstMidEndSepPunct{\mcitedefaultmidpunct}
{\mcitedefaultendpunct}{\mcitedefaultseppunct}\relax
\EndOfBibitem
\bibitem[Chen(1997)]{Chen:1997vq}
Chen,~J.-Y. Development of reduced mechanisms for numerical modelling of
  turbulent combustion. Workshop on Numerical Aspects of Reduction in Chemical
  Kinetics. CERMICS-ENPC Cite Descartes, Champus sur Marne, France, 1997\relax
\mciteBstWouldAddEndPuncttrue
\mciteSetBstMidEndSepPunct{\mcitedefaultmidpunct}
{\mcitedefaultendpunct}{\mcitedefaultseppunct}\relax
\EndOfBibitem
\bibitem[Sung \latin{et~al.}(1998)Sung, Law, and Chen]{Sung:1998gr}
Sung,~C.~J.; Law,~C.~K.; Chen,~J.-Y. An augmented reduced mechanism for methane
  oxidation with comprehensive global parametric validation. \emph{Proc.
  Combust. Inst.} \textbf{1998}, \emph{27}, 295--304\relax
\mciteBstWouldAddEndPuncttrue
\mciteSetBstMidEndSepPunct{\mcitedefaultmidpunct}
{\mcitedefaultendpunct}{\mcitedefaultseppunct}\relax
\EndOfBibitem
\bibitem[Sung \latin{et~al.}(2001)Sung, Law, and Chen]{Sung:2001wa}
Sung,~C.~J.; Law,~C.~K.; Chen,~J.-Y. Augmented reduced mechanisms for {NO}
  emission in methane oxidation. \emph{Combust. Flame} \textbf{2001},
  \emph{125}, 906--919\relax
\mciteBstWouldAddEndPuncttrue
\mciteSetBstMidEndSepPunct{\mcitedefaultmidpunct}
{\mcitedefaultendpunct}{\mcitedefaultseppunct}\relax
\EndOfBibitem
\bibitem[Luo \latin{et~al.}(2012)Luo, Plomer, Lu, Som, and Longman]{Luo:2012cr}
Luo,~Z.; Plomer,~M.; Lu,~T.; Som,~S.~K.; Longman,~D.~E. A reduced mechanism for
  biodiesel surrogates with low temperature chemistry for compression ignition
  engine applications. \emph{Combust. Theor. Model.} \textbf{2012}, \emph{16},
  369--385\relax
\mciteBstWouldAddEndPuncttrue
\mciteSetBstMidEndSepPunct{\mcitedefaultmidpunct}
{\mcitedefaultendpunct}{\mcitedefaultseppunct}\relax
\EndOfBibitem
\bibitem[Curtis \latin{et~al.}(2014)Curtis, Niemeyer, and Sung]{Curtis:2014aa}
Curtis,~N.~J.; Niemeyer,~K.~E.; Sung,~C.~J. An automated target species
  selection method for dynamic adaptive chemistry simulations. \emph{Combust.
  Flame}, in press. \doi{10.1016/j.combustflame.2014.11.004}, 2014\relax
\mciteBstWouldAddEndPuncttrue
\mciteSetBstMidEndSepPunct{\mcitedefaultmidpunct}
{\mcitedefaultendpunct}{\mcitedefaultseppunct}\relax
\EndOfBibitem
\end{mcitethebibliography}
\end{document}